# Additive Manufacturing Approaches for Hydroxyapatite-Reinforced Composites

Mario Milazzo[1, 2]*, Nicola Contessi Negrini[2,3]*, Stefania Scialla[4], Benedetto Marelli[2], Silvia Farè[3], Serena Danti[1,2,5], Markus J. Buehler[2#]

[1] The BioRobotics Institute, Scuola Superiore Sant'Anna, Viale Rinaldo Piaggio 34, 56025 Pontedera (PI), Italy

[2] Dept. of Civil and Environmental Engineering at Massachusetts Institute of Technology, 77 Massachusetts Ave, Cambridge, MA 02139, USA

[3] Dept. of Chemistry, Materials and Chemical Engineering ''G. Natta'', Politecnico Di Milano, P.zza Leonardo Da Vinci 32, Milan, 20133, Italy

[4] Insitute of Nanotechnology (NANOTEC), National Research Council, Via per Monteroni c/o Campus Ecotekne, 73100 Lecce, Italy

[5] Dept. of Civil and Industrial Engineering, University of Pisa, Largo L. Lazzarino 2, 56122 Pisa, Italy

*These authors contributed equally to this work.

#Correspondence should be addressed to:

Prof. Markus J. Buehler

Laboratory for Atomistic and Molecular Mechanics (LAMM)

Department of Civil and Environmental Engineering

Massachusetts Institute of Technology, 77 Massachusetts Ave., Cambridge, Massachusetts 02139 (USA)

mbuehler@mit.edu

+1 617 452-2750

**Abstract:**

Additive manufacturing (AM) techniques have gained interest in the tissue engineering field thanks to their versatility and unique possibilities of producing constructs with complex macroscopic geometries and defined patterns. Recently, composite materials — namely heterogeneous biomaterials identified as continuous phase (matrix) and reinforcement (filler) — have been proposed as inks that can be processed by AM to obtain scaffolds with improved biomimetic and bioactive properties. Significant efforts have been dedicated to hydroxyapatite (HA)-reinforced composites, especially targeting bone tissue engineering, thanks to the chemical similarities of HA with respect to mineral components of native mineralized tissues. Here we review applications of AM techniques to process HA-reinforced composites and biocomposites for the production of scaffolds with biological matrices, including cellular tissues. The primary outcomes of recent investigations in terms of morphological, structural, and *in vitro* and *in vivo* biological properties of the materials are discussed. We classify the approaches based on the nature of the matrices employed to embed the HA reinforcements and produce the tissue substitutes and report a critical discussion on the presented state of the art as well as the future perspectives, to offer a comprehensive picture of the strategies investigated as well as challenges in this emerging field of materiomics.

# LIST OF ABBREVIATIONS

| | | | |
|---|---|---|---|
| 3D | Three-dimensional | NIH3T3 | Fibroblasts from *Mus Musculus* |
| ALP | Alkaline Phosphate | OCN | Osteocalcin |
| AM | Additive Manufacturing | OPN | Osteopontin Gene |
| BMP-7 | Bone Morphogenetic Protein 7 | PA | Polyamide |
| BMSC | Bone Marrow-derived Stem Cells | PCL | Polycaprolactone |
| BSP | Bone Sialoprotein | PE | Polyethylene |
| C3H | Mouse Embryo Cell | PEEK | Poly(ether-ether-ketone) or Poly(oxy-1,4-phenylene-oxy-1,4-phenylenecarbonyl-1,4-phenylene) |
| CH | Chitosan | | |
| COL1 | Collagen I antibody | | |
| CS | Compressive Strength | PGA | Phosphoglyceric Acid |
| CT | Computed Tomography | PLA | Polylactide |
| DGEA | Asp-Gly-Glu-Ala | PLGA | Poly(lactic-co-glycolic acid) |
| DIW | Direct Ink Writing | PLLA | Poly (L-lactic acid) |
| E | Young Modulus | PMMA | Poly(methyl methacrylate) |
| ECM | ExtraCellular Matrix | PVA | Poly(vinyl alcohol) |
| EDC | 1-ethyl-3-3-dimethylaminopropylcarbodiimide hydrochloride | RGD | Arg-Gly-Asp |
| | | SA | Sodium Alginate |
| FDM | Fused Deposition Modeling | SDF1 | Stromal cell-Derived Factor 1 |
| G' | Shear Storage Modulus | SEM | Scanning Electron Microscope |
| G'' | Shear Loss Modulus | SFF | Solid Free Form |
| HA | Hydroxyapatite | SL | Stereolithography |
| HACC | Quaternized Chitosan | SLS | Selective Laser Sintering |
| hASC | human Adipose-Derived Stem Cell | SrHA | Strontium-HA |
| hBMSC | human Bone Marrow Stem Cells | ST-2 | Clonal stromal cell line from bone marrow of BC8 mice |
| HMWPE | High Molecular Weight Polyethylene | | |
| hMSC | human Mesenchymal Stem Cell | T | Temperature |
| h-OB | human Osteoblasts | TCP | Tricalcium Phosphate |
| LAP | Laser Assisted Printing | TE | Tissue Engineering |
| LMWPE | Low Molecular Weight Polyethylene | $T_g$ | Glass Transition Temperature |
| MC3T3-E1 | Osteoblast Precursor Cell Line derived from *Mus Musculus* | TNF | Tumor Necrosis Factor |
| | | $T_{sol-gel}$ | Sol-Gel Temperature |
| MD | Molecular Dynamics | UHMWPE | Ultra-High Molecular Weight Polyethylene |
| MG-63 | Osteosarcoma cell | UV | Ultra Violet |
| MPP | Mitochondrial Processing Peptidase | η* | Complex Viscosity |
| MSC | Mesenchymal Stem Cell | MPP | Mitochondrial Processing Peptidase |

**Introduction**

Since 1984, when Prof. Charles Hull showed one of its first uses at the Massachusetts Institute of Technology, Additive Manufacturing (AM) has been widely employed in many fields, such as the arts, food industry, manufacturing, and design. However, AM found a remarkable framework in 1993, when a team of researchers published a study concerning its employment within the biomaterials and Tissue Engineering (TE) field.[1] These researchers explored for the first time the capabilities of this technique in the TE field as a means to repair tissue defects and damaged organs.[2] This underlying idea consisted of creating resorbable 3D-structures, referred as scaffolds, to regenerate tissues. AM, as largely demonstrated, has allowed scaffolds to be tailored in terms of specific features such as shape, micro/macro porosity, and pore interconnectivity ratio while, at the same time, the fabrication time was significantly reduced.[3] Since then, AM has spread into a wide family of techniques to fabricate customized and complex three-dimensional (3D) constructs, being a much more versatile alternative to other traditional approaches where complex topologies are difficult to achieve (e.g., electrospinning,[4] machining,[5] screw extrusion,[6] solvent casting/particulate leaching,[7], freeze-drying,[8], gas foaming,[9] replica molding and masking).[10] Although the here cited conventional fabrication techniques represent important milestones for HA-reinforced composites production for TE, they achieve a controlled internal/external architecture of the scaffolds (i.e., dimensions, pore morphology, pore size, pore interconnectivity) only in a limited manner. Additionally, some of these conventional techniques are time-consuming, since they need to use molds and manual control. Consequently, these fabrication processes are less intensive and, sometimes, poorly reproducible, compared to AM technologies, which can represent a more efficient tool to produce personalized HA-reinforced composites with high flexibility in terms of shape and size, high reproducibility in a relatively short time, thus overcoming the drawbacks of traditional manufacturing processes.

These 3D scaffolds have then been extensively explored as supports for cell attachment, growth and differentiation, aiming at forming new extracellular matrix,[11,12] promoting *in vivo* tissue regeneration or fabricating *in vitro* tissue models.

Despite a number of mere materials can be processed by AM to obtain scaffolds, an increasing interest has been focused on processing composites, multiphase materials able to maximize the properties of the single components, when taken alone.[13] Composites for TE are usually composed of a matrix, a reinforced part (i.e., particles, fibrils and flakes) and, possibly, cellular tissues. Composites used for AM purposes (i.e., biomaterial inks or bioinks)[14] can be classified in two groups: hard and soft materials, depending on the matrix used for each specific application.[15–18] The first class includes materials with matrices possessing mechanical properties that assure stability after the printing (e.g., metals, ceramics, and many thermoplastic polymers). However, due to the procedures needed for processing such materials, it has been shown that cells must be seeded at completed fabrication process of the scaffold to assure a sufficient biological viability. As for the second class, soft matrices are typically hydrogels, hydrophilic polymers able to hold a large amount of water (from 10 - 20% to thousands of times their dry weight)[19] while assuring structural integrity when adequately processed,[20,21] and allowing permeability to nutrients to promote new tissue formation.[22–24] Figure 1 summarizes the different matrices employed to fabricate HA-reinforced composites that will be reviewed in this paper.

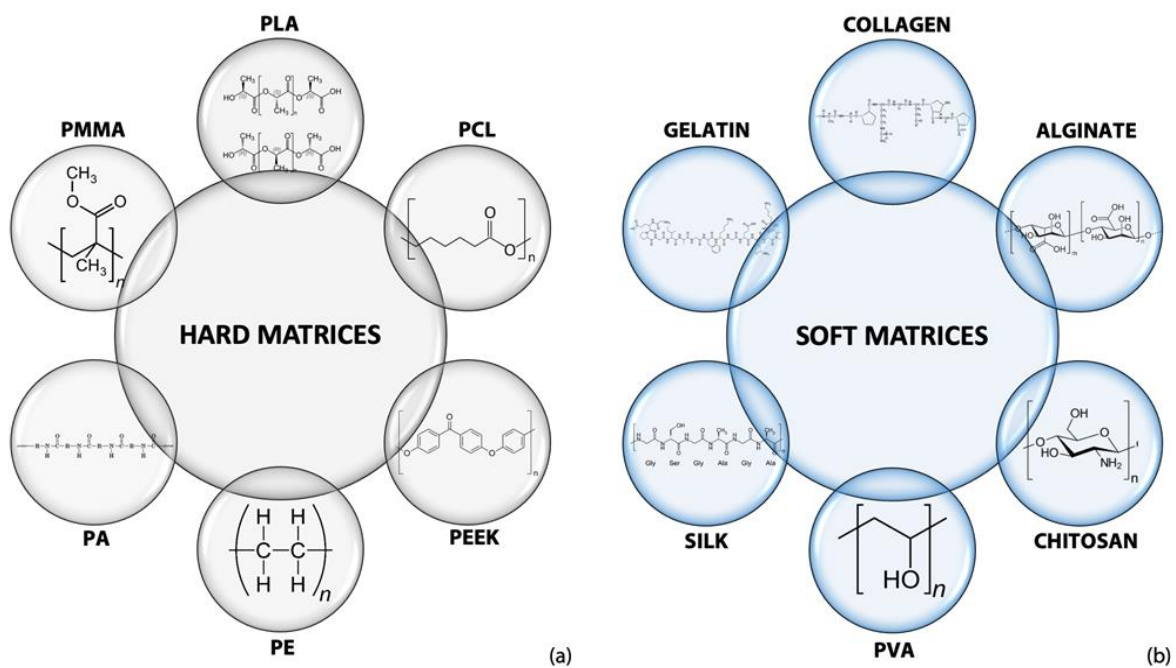

*Figure 1. (a) Hard and (b) soft HA-reinforced matrices fabricated via additive manufacturing for tissue engineering.*

Besides their hard/soft nature, there are specific features that each material must comply with in order to be a good candidate for the fabrication of scaffolds by AM. Firstly, biomaterial inks must possess rheological properties that allow their processability. Since this feature plays a dominant role in the selection of the AM technique, several studies have been published to find the optimized rheological features for specific material classes (e.g., viscosity, extensional viscosity, surface tension depending on the printing technique).[25] Furthermore, the mechanical properties must be suitable for the targeted application (e.g., bone reconstruction);[26] this issue is particularly relevant for soft biocomposites where a crosslinking process is often pursued to overcome this concern. The last structural property to be taken into account is the capability of the scaffold in furnishing an adequate porosity: although several authors have achieved good results in making unidirectional porosities,[27] a multidirectional scenario (i.e., interconnections between pores) may help over a more homogenous tissue infiltration and vascularization, features that prevent necrosis and rejections in implant sites.[28]

From a biological point of view, the composites must be a non-toxic environment for efficient cell seeding, proliferation, and differentiation, possibly granting progressive *in situ* integration within the host tissue. As exhaustively detailed in the following sections, this property is significantly connected to the AM techniques employed to process the composites.

The aim of this paper is to review the advancements for HA-reinforced materials by proposing a classification based on the employed matrices (i.e., hard and soft) in the framework of the AM techniques. Finally, a critical discussion on current and future challenges in the field is treated.

**Additive manufacturing techniques for HA and HA-reinforced composites**

Tissue regeneration by HA-reinforced tissue-engineered scaffolds processed by AM requires three fundamental steps: (1) identification of the missing/damaged tissue and generation of its digital topology (Figure 2A), (2) production of the composite and fabrication of the scaffold/substitute via AM (Figure 2B), and (3) replacement of the native tissues (Figure 2C). Besides, the fabrication step is crucial, and it plays a crucial role in the process, critically affecting the final outcome of the tissue regeneration.

According to the American Chemical Society, it is possible to roughly classify AM methods for tissue engineering in three main categories: extrusion-based, laser-based and droplet-based techniques.[29] Due to the high viscosity of the HA-reinforced materials given by the addition of the ceramic component, droplet-based techniques cannot be employed since they are compatible only with low viscosities (< 10 mPa·s) and high shear rates ($10^5 – 10^6$ s$^{-1}$).[30–35] In contrast, HA-reinforced composites have been successfully fabricated by extrusion-/laser-based processes (Figure 2B).

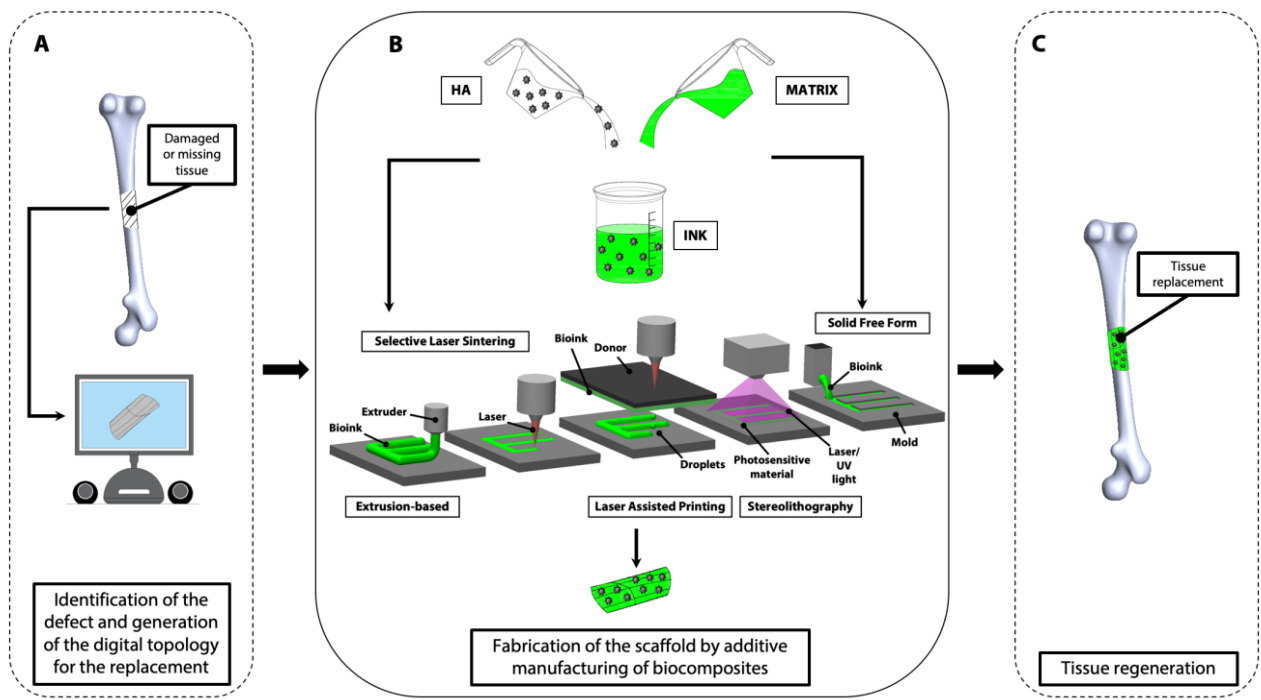

*Figure 2. Additive manufacturing techniques used for the fabrication of hydroxyapatite-reinforced composites as scaffolds for tissue engineering applications. From (A) the generation of the digital topology of the damaged/missing tissue to (B) the fabrication of the scaffold through AM approaches with a HA-based bioink and (C) the final tissue regeneration.*

For extrusion-based processes, including Direct Ink Writing (DIW, also called Robocasting) and Fused Deposition Modeling (FDM), the raw material is ejected by a nozzle, namely an extruder, steered through a mechanical or electromagnetic actuator. While in FDM the material, usually synthetic thermoplastics, is printed at high temperatures (140 – 250 °C) at melt state to reduce the viscosity,[36] DIW allows the extrusion at low temperatures (e.g., room temperature).[16] This specific advantage of DIW makes it suitable also to print cells without affecting their viability, not possible when using FDM. It has been demonstrated that suitable materials for DIW/FDM possess specific ranges of viscosity (i.e., from 30 to $6 \times 10^7$ mPa·s): higher-viscosity materials are generally used for structural parts whilst lower-ones are usually employed for building a proper environment for cell viability.[37]

Laser-Based techniques include Stereolithography (SL), Selective Laser Sintering (SLS) and Laser Assisted Printing (LAP). SL is the oldest approach and consists of the projection of a direct-light (i.e., Ultra Violet (UV), laser) on a viscous photosensitive polymer solution to crosslink it and, thus, promote its solidification. Alternatively, with SLS, a laser beam is focused on the material, in the shape of powder, to selectively sinter it.[38] From a biological point of view, SLS is unsuitable for cell encapsulation due to the high energy density employed to make the construct; contrary to SLS, under certain conditions, SL assures high cell viability.[39–41] Finally, in LAP, a laser (spot size = 40 – 100 µm) is concentrated on a donor layer that buffers the energy supply to the biomaterial ink, eventually deposited on a receiver surface in the shape of droplets. The interposition of an absorbent thick layer also allows the direct printing of living tissues (i.e., cells) without significantly affecting their viability.[42–44] Finally, in contrast to the extrusion-based techniques, laser-based approaches can process materials with a range of viscosity (i.e., 1-300 mPa·s).[45]

A different approach, often referred to as indirect AM, is represented by the Solid Free Form technique (SFF). Historically, the SFF term has been associated as a synonym of AM, hence it a family of techniques that aims at fabricating structures with a layer-by-layer approach from an image-based/3D topology.[46] In this review, however, the term SFF will be referred to the cases in which AM is used to create molds in which to cast a slurry of material to be eventually treated (e.g., dehydration) to enhance the crosslinking and tune pore interconnectivity.[47] The evident simplicity of the approach is, at the same time, its main drawback since the topology of the shapes intrinsically sets the constraints on the design of the constructs.

**Hydroxyapatite**

HA is a ceramic material with a hexagonal crystallographic structure, chemically described by the $Ca_{10}(PO_4)_6(OH)_2$ formula (Figure 3). Its successful employment in TE is due to its chemical structure, similar to the mineralized constituents of bone. Although native tissues do not contain neat HA since they possess other impurities (e.g., carbon, hydrogen phosphate, etc.), HA has been commonly used for bone regeneration purposes[47–50] due to its excellent physicochemical properties such as osteoconductivity, bioactivity, resorbability and slow-decaying properties.[51–54] One of the most interesting features of HA concerns the cell response modification based on its dimensions: nanometer sizes increase intracellular uptake and reduce cell viability *in vitro*.[55] Moreover, size and crystallinity of HA particles may affect stability and inflammatory response, increased if the dimensions are smaller than 1 μm.[56]

Also from a mechanical point of view, HA shows interesting features in terms of stiffness (Young Modulus E = 35 – 120 GPa) and compressive strength (CS) (120 – 900 MPa)[57] and toughness investigated by a number of molecular dynamics (MD) models and confirmed experimentally.[58–60] For instance, microspheres of HA have been studied by many research teams for bioengineering applications due to their potential as local drug and protein delivery systems.[56,61–71] On the other hand, few studies have investigated the incorporation of HA microspheres within polymeric matrices produced via AM.[63,72]

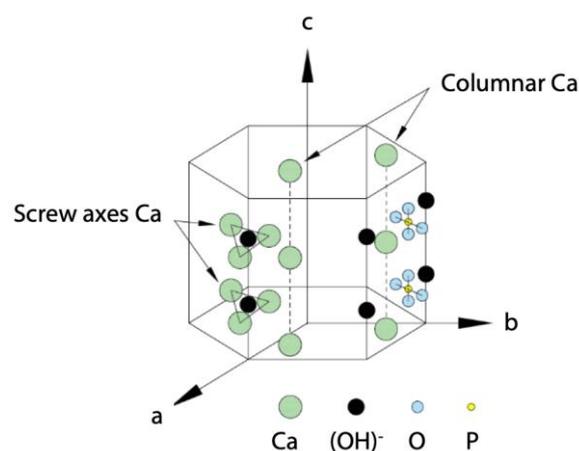

*Figure 3. Hexagonal crystallographic structure of the hydroxyapatite crystal.*

Researchers have been fabricating neat HA structures by means of different AM techniques building osteoconductive scaffolds with tuned microstructures with pores ranging from 100 to 1200 μm,[73–75] able to successfully support the formation of new vascularized tissues.[76] Different approaches have been pursued: SL,[77] LAP,[78] SFF,[74,75,79] SLS [73] and DIW.[53,80,81] Whereas with laser-based techniques it was possible to obtain structures that, tested both *in vitro* and *in vivo* in rodent or canine models, demonstrated good adhesion, proliferation, and osteochondral differentiation of the seeded human Mesenchymal Stem Cells (hMSCs) and absence of inflammation,[73,77,78] it is with SFF and DIW that the best results were achieved.

SFF is the simplest method involving AM only for preparing molds in which to cast a slurry of material composed of HA powder mixed with demineralized water and chemicals. Despite its intrinsic simplicity, it was possible to get stratified structures – layer thickness in the order of 0.05 mm with higher porosity (i.e., up to 44%) in contrast to DIW (i.e., up to 37%).[75,79]

DIW of HA is generally hard to perform without intermediate steps aimed at treating raw materials characterized by low crystal growth rates (i.e., 2.7 x $10^{-7}$ mol $Ca_{10}(PO_4)_6(OH)_2$ $min^{-1} \cdot m^{-2}$).[80,81] Thus, Leukers *et al.* fabricated scaffolds made of spray-dried HA-granulate with polymeric additives to improve bonding and flowability and a polymer blend (Schelofix) as binder. In order to improve structural strength, a further sintering was performed at 1300 °C for 2 h revealing, eventually, how the seeded cells grew especially among the granules.[53]

Miranda *et al.* employed DIW with a 250 μm nozzle to deposit HA filaments at 20 mm·$s^{-1}$. After printing, the constructs were dried at 400 °C for 1 h to allow the evaporation of the organics, achieving a porosity close to 40% with Young Modulus equal to 7 ± 2 GPa.[82] Furthermore, Carrel *et al.* prepared stratified HA and tricalcium phosphate (TCP) scaffolds to evaluate bone regeneration in sheep calvarian model. Compared to standard granular substitutes, HA structures showed higher osteoconductivity properties.

As discussed above, 3D printed HA scaffolds have been widely employed in tissue engineering to regenerate/substitute bone tissue (Table 1). Due to its intrinsic stability and shapeability, HA-based scaffolds have been successfully fabricated with all the previously listed AM techniques, with the only exception represented by the droplet-based technique, probably due to intrinsic high viscosity that may induce clots in the cartridges. Scaffolds with tunable pore size (300 - 1250 μm) for hosting viable cells were successfully fabricated. As reported in Table 1, biological *in vitro* assessments demonstrated how HA-based constructs are able to promote cell proliferation and differentiation. Moreover, a few number of *in vivo* tests on animals have shown how implants might induced mild inflammation reactions close to the implantation sites that, however, were completely resorbed in less than one month.[78]

Despite the successful validation of pure HA-based scaffolds, this ceramic component has also gained interest as reinforcement for composites processed via AM.

*Table 1. Literature review of HA constructs processed by additive manufacturing technologies.*

| AM method | Porosity [%]/ Pore size [μm] | Structural properties | Cell type | Cellular response | Reference |
|---|---|---|---|---|---|
| SL | 79.6% / 1250 μm - 69.3% / 790 μm - 48.2% / 500 μm | - | hMSC | Cell proliferation, and osteochondral differentiation | [77] |
| LAP | - | - | None – *In vivo* laser printing | Inflammation that was completely resorbed after 21 days | [78] |
| SFF | 44% - 300 μm | | Clonal stromal cell line from bone marrow of BC8 mice (ST-2) | Cell proliferation and differentiation | [75] |
| SFF | 400 - 1200 μm | | - | - | [74] |
| SFF | 52% / 286 - 376 μm | | Goat Bone marrow-derived stem cells (BMSC) | Cell proliferation and ectopic bone formation | [79] |
| SLS | - | Bending strength: 66.2 MPa | - | - | [73] |
| DIW | 42% | CS = 27 MPa | - | - | [81] |
| DIW | 500 μm | - | Osteoblast Precursor Cell Line derived from Mus musculus (MC3T3-E1) | Cell proliferation and growth | [53] |
| DIW | - | - | human osteoblast cells (h-OB) | Cell spreading, adhesion and differentiation | [80] |

**Hydroxyapatite-reinforced biocomposites via additive manufacturing**

A possible classification for composites is based on the matrix used for the constructs. In the specific case of the HA-reinforced biomaterials, only few materials have been processed via AM (Table 2).

*Table 2. Literature review of hydroxyapatite-reinforced composites processed by additive manufacturing technologies for the production of scaffolds.*

| AM techniques | EB | SLS | SFF |
|---|---|---|---|

| | | Material | | | |
|---|---|---|---|---|---|
| **Composites** | **Hard Matrices** | Polylactide (PLA) | [83–86] | - | [72] |
| | | Polycaprolactone (PCL) | [87–91] | [92,93] | - |
| | | Poly(ether-ether-ketone) (PEEK) | - | [94] | - |
| | | Polyethylene (PE) | - | [95] | - |
| | | Polyamide (PA) | [96] | - | [97] |
| | | Poly(methyl methacrylate) (PMMA) | - | [98–100] | - |
| | **Soft Matrices** | Alginate | [101–106] | - | - |
| | | Collagen | [107] | - | [108,109] |
| | | Gelatin | [110–113] | - | - |
| | | Chitosan (CH) | [114–119] | - | - |
| | | Poly(vinyl alcohol) (PVA) | - | [120] | - |
| | | Silk | [25,121] | - | |

The following two sections will present the main achievements over fabricating HA-reinforced composites via AM, based on the classification of the polymeric matrices (Figure 1).

**Hard matrix-based composites**

Hard matrices employed so far for HA-reinforced composites are synthetic polymers (e.g., PLA, PCL). Table 3 reports the main aspects for each study, detailed also in the following sections, focusing on the structural properties and biological responses.

*Table 3. Hard matrix-based hydroxyapatite reinforced composites: structural, morphological and biological properties.*

| Material | Pore size [µm]/ porosity [%] | AM method | Structural properties | HA size/ distribution | Cell type | Cellular response | Reference |
|---|---|---|---|---|---|---|---|
| PLA | 55% | DIW | E in parallel to the printing plane: 150 ± 40 MPa; | 70 wt% | - | - | [90] |
| | | | E in perpendicular to the printing plane: 84 ± 9 MPa | | | | |
| | 500 μm | SFF | CS = 1.46 MPa | Average diameter 10 μm CH/HA 50:50 and CH/HA 60:40 | MC3T3-E1 pre-osteoblastic cells | Osteoblastic cells proliferation and differentiation | [72] |
| | 300 μm/ 76% | DIW | CS = 30 MPa E = 1.9 GPa | - | hBMSCs | Adhesion, proliferation and differentiation | [84,85] |
| | - | DIW | CS = 2.443 ± 0.514 MPa | 3 wt% HA with 97 wt% PLA | NIH3T3 fibroblasts from *Mus Musculus* | Bone regeneration | [86] |
| | - | DIW | CS = 2.446 ± 0.467 MPa | 3 wt% HA with 94 wt% PLA and 3 wt% Silk | NIH3T3 fibroblasts from *Mus Musculus* | Bone regeneration | [86] |
| PCL | 26% | FDM | CS = 15 MPa E = 80 MPa | 30 wt% HA | MC3T3-E1 | good cell biocompatibility, and biodegradation ability | [88]. |
| | 200 μm | DIW | - | 20 wt% HA | Grow factors: Stromal cell-derived factor 1 (SDF1) and Bone morphogenetic protein 7 (BMP7) | Bone regeneration and periodontal integration | [87] |
| | 500 μm | DIW | - | 0.03 μm / HA (4 mg):PCL (700 mg) | Mesenchymal Stem Cell (MSCs) | Excellent osteoconductive and osteointegration properties. High histocompatibility | [89] |
| | 55% | DIW | E In parallel to the printing plane: 110 ± 20 MPa | 70 wt% | - | - | [90] |

| | | | E in perpendicular to the printing plane: 24 ± 5 MPa | | | | |
|---|---|---|---|---|---|---|---|
| | 37% | SLS | E = 67 MPa<br>CS = 3.2 MPa | HA particles diameter 10 – 100 μm | BMP-7 | Tissue in-growth | [92] |
| | 410 μm | SLS | E = 2.3 MPa<br>CS = 0.6 MPa | 30% wt% of HA | - | - | [93] |
| | - | DIW | CS = 7 MPa<br>E = 40 MPa | Strontium-HA:PCL [0:100, 10:90, 30:70, 50:50 wt%] | BMSCs | Cell proliferation and osteogenic differentiation. Promotion of bone regeneration | [91] |
| PEEK | - | SLS | - | Powder Diameter below 60 μm/ 10 – 40 wt% | - | - | [94] |
| PE | 200 – 400 μm | SLS | - | HA 30% - 40% wt% | - | - | [95] |
| PA | 300 – 500 μm | SFF | CS = 117 MPa<br>E = 5.6 GPa | - | - | - | [97] |
| PA | - | DIW | - | - | - | - | [96] |
| PMMA | 40 -100 μm / 30% | SLS | - | 50 g – 100 ml $H_3PO_4$ 14 wt% PMMA | - | - | [99,100] |

**Polylactide (PLA)**

Polylactide or polylactic acid (PLA) is among the most widely bioresorbable polymers approved by the U.S. Food and Drug Administration (FDA) for human clinical applications such as surgical sutures, cranio-maxillofacial augmentation, bone fixation, soft-tissue implants and implantable scaffolds.[122] It is a biocompatible polymer, synthesized by multiple ways (e.g., poly-condensation, ring-opening or enzymatic polymerization) from lactic acid monomers, which exist as two enantiomers, L- and D-lactic acid. This confers chiral property to PLA, for which it exists in different stereoisomers: poly(L-lactide) (PLLA), poly(D-lactide) (PDLA), and poly(DL-lactide) (PDLLA). The biocompatibility of this polymer makes it highly attractive for biomedical applications, in particular because its degradation products do not interfere with the healing process but, rather, they are metabolized through Krebs cycle and expelled.

The chirality of monomer's unit influence considerably its physico-chemical, mechanical and rheological properties, and consequently its degradation rate. In particular, by varying the enantiomeric units' content, the PLA's crystallinity may be controlled ranging from an amorphous/semi-crystalline to crystalline, increasing the mechanical feature of the polymer. Semi-crystalline PLA has an approximate tensile modulus of 3 GPa, tensile strength of 50 – 70 MPa, flexural modulus of 5 GPa, flexural strength of 100 MPa, and an elongation at break of about 4%.[123] Also the molecular weight ($M_W$) has a significant impact on the mechanical properties and degradation rate of the PLA. High $M_W$ PLA has a very high resorption time (2 - 8 years) that *in vivo* might lead to inflammation and infection.[124,125] Therefore, production of low $M_W$ PLA is desirable because it provides a shorter degradation rate, but at the same time it is able to guarantee suitable mechanical features for a period of time, which fit with bone fracture healing.[123] PLA with $M_W$ lower than 2000 Da behaves like a hydrogel, useful for drug delivery applications. From a mechanical point of view, it has been reported and increasing of the tensile modulus of PLLA of 2 time when the $M_W$ is raised from 50 to 100 kDa,[123,126] and tensile strengths of 15.5, 80 and 150 MPa, for varying the $M_W$ from 50 over 150 to 200 kDa, respectively.[123,127]

Nevertheless, PLA processing parameters (such as injection or extrusion with or without an annealing step) affect the mechanical behavior of the polymer (Table 4). More specifically, a fast cooling after injection molding did not allow the polymer chains to rearrange into a crystalline structure; while after annealing, an important crystalline structure was recreated. PLA is also relatively hydrophobic, and this induces a lower cell affinity and lead to an inflammatory response from the living host upon direct contact with biological fluids. Moreover, also the lack of reactive side-chain groups makes PLA chemically inert, and consequently its surface and bulk modifications are among the main research topics.[123]

*Table 4. Mechanical properties (E, yield strength and elongation at break) of PLA processed by injection (PLA-I) and extrusion/injection (PLA-EI) without or with annealing (PLA-EIA).*[123]

| Mechanical parameters | PLA_I | PLA_EI | PLA_IA | PLA_EIA |
|---|---|---|---|---|
| **E [GPa]** | 3.7 ± 0.1 | 3.9 ± 0.1 | 4.1 ± 0.1 | 4.1 ± 0.1 |
| **Tensile strength [MPa]** | 65.6 ± 1.3 | 65.2 ± 0.9 | 75.4 ± 0.9 | 77.0 ± 1.1 |
| **Elongation at break [%]** | 4.0 ± 0.8 | 5.4 ± 0.6 | 2.5 ± 0.2 | 3.3 ± 0.3 |

PLA-based constructs reinforced with HA powder (diameter ranging 1 - 3 μm) were successfully printed via DIW by Russias *et al.*, getting structures with acceptable porosity – about 55%. The ink was prepared by mixing 70 wt% of HA powder with PLA polymer and ethanol to tune the viscosity of the compound. The printing nozzle (diameter = 5 - 410 μm) ejected the slurry with a deposition speed ranging between 5 – 20 mm·s$^{-1}$: it was observed that lower speeds lead to clotting inside the nozzle whilst faster ones make the slurry discontinuous. Although no specific biological tests were performed, a mechanical assessment was carried out by evaluating the response of the constructs in parallel or perpendicularly to the printing plane, achieving results statistically comparable, but still far from the requirements needed to replace the bone: 150 ± 40 MPa and 84 ± 9 MPa in parallel and perpendicularly to the printing plane respectively (Figure 4).[90] Li *et al.* developed scaffolds by using SFF to fabricate a three-component scaffold with a woodpile structure consisting of PLLA, chitosan (CH) and HA microspheres at different ratios (60:40 and 50:50 CH/HA), with a macro-porosity of more than 50% together with micro-pores induced by lyophilization. The mechanical properties of the PLLA/CH/HA composite scaffolds were compared with that of CH/HA 50:50 and CH/HA 60:40 composites, obtaining 1.46, 1.31 and 1.07 MPa as compression modulus, respectively, which suggest the use of these scaffolds for non-load-bearing bone implants.

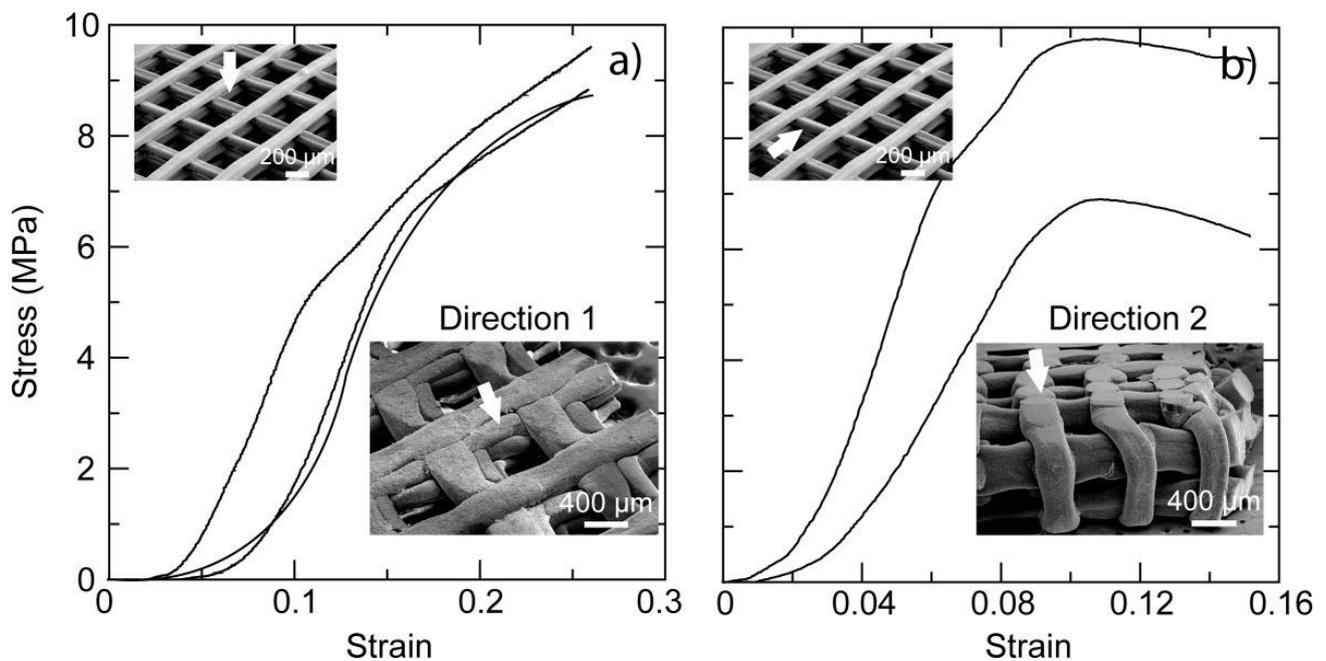

*Figure 4. Stress-strain curves from compression tests made on PLA/HA scaffolds (70 wt% HA) along the (a) perpendicular and (b) parallel directions with respect to the printing plane. The slopes, namely E, resulted equal to (a) 84 ± 9 MPa and (b) 150 ± 40 MPa. Lenses shows Scanning Electrode Microscope (SEM) micrographs after the mechanical test. Reprinted from [90] with permission – © 2017 John Wiley and Sons.*

Comparing this with the data for human bone, the CS of the composite scaffold is still far from that of cortical bone (CS = 130 – 180 MPa, E = 12 – 18 x $10^3$ MPa) and cancellous bone (CS = 4 – 12 MPa, E = 0.1 – 0.5·$10^3$ MPa), but is closer to cartilage (strength of 4 – 59 MPa and modulus of 1.9 – 14.4 MPa).[72,128]

These scaffolds showed excellent biocompatibility and ability for three-dimensional tissue growth formed by MC3T3-E1 pre-osteoblastic cells. The pre-osteoblastic cells cultured on these scaffolds formed a network on the HA microspheres and proliferated not only in the macro-pore channels but also in the micro-pores themselves. The presence of PLLA in the composite scaffolds improved the initial cell proliferation and differentiation process up to 4 weeks, as revealed by the tissue growth and Alkaline Phosphate (ALP) enzyme activity. In the later stages, at 5 weeks, a decrease in ALP was observed for PLLA composite scaffolds which might be due to a partial degradation of the polymer.[72] Alternatively, other authors[86] added silk fibroin nanoparticles to a HA-PLA composite filaments used to 3D print bone clips to promote bone regeneration. The authors demonstrated that, after achieving an accurate 3D printed device obtained by computer tomography (CT) of a rat femur, the silk-HA-PLA composite showed adequate mechanical properties and enhanced bone regeneration when compared to the HA-PLA composite without silk.

Later, to confer antimicrobial properties to the biomaterials, Yang *et al.* [84,85] used a quaternized chitosan (hydroxypropyltrimethylammonium chloride - HACC) grafted to a FDM-printed HA-reinforced poly(lactic-co-glycolic acid) – PLGA - composite scaffold manufactured layer-by-layer (up to 24 layers).[84,85] Quaternized CH consists in the introduction of permanent positively charged quaternary ammonium groups to the polymer to enhance its water solubility and antibacterial feature over a broad pH range.[129] The 3D-printed HACC/PLGA/HA scaffolds showed a large homogeneous macro-porosity of about 76%, with an average pore size of about 300 µm and a highly interconnected micro-porosity, which could provide a suitable substrate for cell infiltration and bone ingrowth. The addition of HA induced an improvement of mechanical properties as demonstrated by the CS = 30 MPa and E = 1.9 GPa, which appeared to be intermediate between mechanical properties of the cancellous and cortical bone.[130] These results demonstrated that this approach allowed to design and manufacture 3D-printed scaffolds as bone substitutes with mechanical properties reassembling closely those of bone tissues, which could be applied *in vivo* for cortical and cancellous bone repair.[85] The authors also proved that the use of quaternized chitosan decreased bacterial adhesion and biofilm formation under *in vitro* and *in vivo* conditions, disrupting microbial membranes and inhibiting the biofilm formation. In addition to the antibacterial activities of these 3D-printed scaffolds, attention was also paid to their biocompatibility and osteogenic activity. They demonstrated, in fact, that the incorporation of HA into the scaffolds significantly improved human Bone Marrow Stromal Cells (hBMSCs) adhesion, proliferation, spreading and the expressions of several critical osteogenic differentiation-related genes: early osteogenic markers Bone Sialoprotein (BSP) and Collagen I antibody (COL1), and relatively late markers Osteocalcin (OCN) and Osteopontin (OPN) genes. Furthermore, they also proved that HACC/PLGA/HA scaffold provided a satisfactory *in vivo* micro-environment for the neovascularization and tissue integration, which laid good foundation for the regeneration of bone defects *in situ*.

**Polycaprolactone (PCL)**

Polycaprolactone (PCL) is an aliphatic thermoplastic polyester prepared by the ring-opening polymerization of the cyclic monomer ε-caprolactone,[131] which can proceed via anionic, cationic, coordination or radical polymerization mechanisms.[132] PCL is a hydrophobic, semi-crystalline polymer with a very low glass-transition temperature of ($T_g$ = -60 °C); thus, it is commonly in the rubbery state with a high permeability under physiological conditions.[122] It is biodegradable but more stable compared to PLA because of its less frequent ester bonds per monomer; therefore, degradation takes longer, from several months to several years, for PCL chain fragments to be degraded in the body. This depends on its molecular weight, degree of crystallinity, and the conditions of degradation.[131] Its rubbery features, tailorable degradation kinetics and mechanical properties, ease of shaping and manufacturing have made PCL, over time, a suitable material to fabricate scaffolds for replacing hard tissues, an interesting material for surgical sutures, and micro- and nano-drug delivery.[133] The biocompatibility of PCL has been proved by non-toxic and non-acidification effects in *in vivo* experiments.[134–136] Structurally, despite its low E in the range of 0.21 – 0.44 GPa,[127] it has been often used for bone and cartilage grafting and repair being prepared via AM.[137–140]

PCL/HA reinforced scaffolds have been fabricated by employing three different additive manufacturing techniques: FDM, DIW and SLS. FDM was used to create a replacement for goat femurs by using 30 wt% HA, getting a 26% porous structure with 15 MPa of CS and E of 80 MPa. *In vitro* studies showed that the samples seeded with MC3T3-E1 proved good cytocompatibility and successful biodegradation.[88]

Concerning the DIW approach, a first study described a Poly-ε-caprolactone reinforced with HA (20 wt%) 3D-printed to mimic human molars and rat incisors. *In vivo* experiments in rodents showed new native alveolar bone grown on the implant site.[87] Meseguer-Olmo *et al.* presented, instead, a scaffold made of PCL with nano-crystalline silicon-substitute HA (nano-SiHA) with demineralized bone matrix. They employed DIW to deposit a polymerized slurry. Before testing the samples *in vivo* on rabbits, the construct was dried at 37 °C and sterilized by UV. The post-implant assessment showed excellent osteoconductive and osteointegration properties along with high histocompatibility with the host tissues.[89] Alternatively, Russias *et al.* focused their study on the mechanical properties of PCL/HA constructs (HA 70 wt%) made by DIW. Results revealed how the E in parallel to the printing plane (i.e., E = 110 ± 20 MPa) differed from that in perpendicular it (i.e., 24 ± 5 MPa) without further biological investigations (Figure 5).[90]

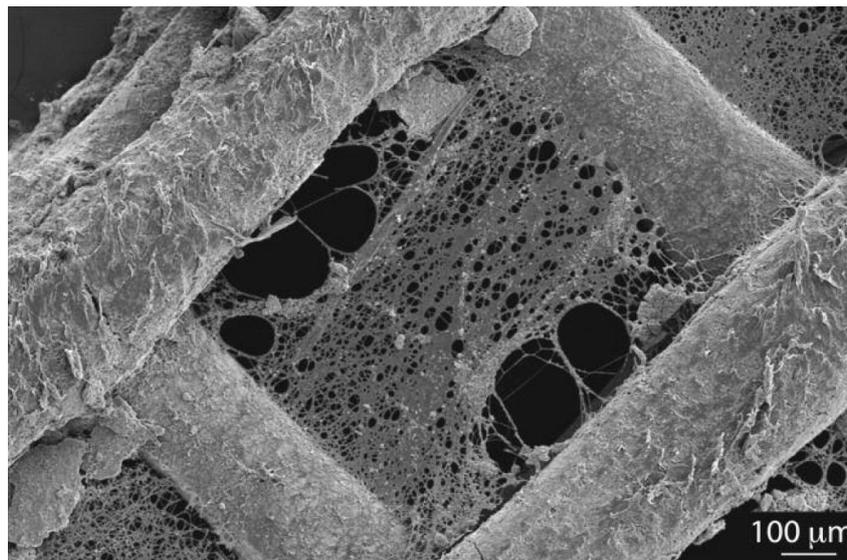

*Figure 5. SEM micrograph of a PCLA/HA (HA 70 wt%) scaffold where the surface degradation is appreciable. Reprinted from [90] with permission – © 2017 John Wiley and Sons.*

SLS was exploited to fabricate replacements able to promote tissue in-growth: by using PCL/HA composites, Williams *et al.* fabricated a porous scaffold (37% porosity) able to perform the envisaged task although possessing relatively poor mechanical properties (i.e., E = 67 MPa, CS = 3.2 MPa).[92] Weaker results were also achieved in another work (i.e., E = 2.3 MPa, CS = 0.6 MPa) although they used HA 30 wt%.[93] More recently, by employing Strontium-HA (SrHA) compounds, Liu *et al.* prepared 3D printed scaffolds with different percentages of reinforcement (0:100 to 50:50 wt%) to test *in vitro* and *in vivo* the capability of the constructs to promote bone regeneration. The samples showed a compressive stress in the order of 7 MPa with E of 40 MPa with fair biological outcomes revealing the proliferation and the osteogenic differentiation of the BMSCs seeded into them.[91]

**Poly(ether-ether-ketone) (PEEK)**

Poly(ether-ether-ketone) or Poly(oxy-1,4-phenylene-oxy-1,4-phenylenecarbonyl-1,4-phenylene) is a polyaromatic semi-crystalline thermoplastic polymer, also known as PEEK, which has been increasingly employed as material for prosthetics, orthopedics, maxillo-facial and spinal implants. It represents an efficient alternative to implantable metal-based devices, due to its versatile mechanical and chemical properties that are retained at high temperature, reducing shielding stress after implantation: E in the range of 3 – 4 GPa,[141] close to that of the human cortical bone (i.e., 7 – 30 GPa),[142] and a tensile strength of 90 – 100 MPa.[143] Moreover, it has also radiolucent properties, permitting radiographic assessment.

PEEK-based materials have been considered relatively bio-inert in biological environments, demonstrating a weak osteointegration following implantation. For this reason, over the past decade, there has been a growing interest in further improving PEEK features to stimulate bone apposition for load-bearing orthopedic applications.[144] Scaffolds made of a blend composed of HA and PEEK have been fabricated via SLS: the authors performed a sensitivity analysis on the main parameters of the SLS (e.g., laser power) along with different percentages of HA powder, from 10 to 40 wt%, with diameter size distribution below 60 μm. The optimized constructs were fabricated by fixing the laser at 16 W and 140 °C. However, they decided to avoid higher HA percentages (i.e., more than 40 wt%) to avoid instability and fragility of the final structure.[94]

**Polyethylene (PE)**

Polyethylene (PE) is a versatile thermoplastic polymer largely employed in the orthopedic field, as a load-bearing mean in artificial joints or to treat arthritis.[145,146] It is an inert and hydrophobic material that does not degrade *in vivo*. PE is produced at different molecular weights ($M_w$) and different crystallinity grades. Based on the $M_w$, PE is classified as low-molecular weight polyethylene (LMWPE), high-molecular weight polyethylene (HMWPE), and ultrahigh-molecular weight polyethylene (UHMWPE). LMWPE with $M_w$ 50 – 200 kDa and 40 – 50% crystallinity is the softest, with E equal to 100 – 500 MPa, mainly used for packaging applications. HMWPE can have similar $M_w$ but 60 – 80% crystallinity and E of 400 - 1500 MPa. It has been used to make stable devices (i.e., containers) or for implantations. UHMWPE has $M_w$ above 2000 kDa, 50% – 60% crystallinity and E of 1000 – 2000 MPa.[145] In addition to their $M_w$, HMWPE and UHMWPE have gained much attention as load-bearing materials for joint endoprostheses due to their chemical inertness, mechanical strength, limited tissue reaction, and biostability.[146] Furthermore, it was observed that it is possible to tune the wear resistance in UHMWPE by varying the grade of crosslinking, in order to make the material more suitable for specific applications.[147]

To enhance the biocompatibility of PE, researchers have explored the use of PE as a matrix with HA, developing composites with improved features. Most of the PE/HA composites have been fabricated via extrusion technologies, non-AM driven, without any specific control of the macro-topology.[148–155] Alternatively, Hao *et al.* investigated the development of a structure made of HDPE reinforced with HA (at 30 – 40 vol%) made by a $CO_2$ SLS (spot = 193 μm, wavelength = 10.6 μm, focal length = 491 mm). They highlighted how this specific AM process plays a role in the key features of the composite, affecting the fusion of HA particles, enhancing their bioactivity, and finely tuning the pores size, reaching 200 – 400 μm pores with an optimized laser power of 2.4 W at 1200 mm·s$^{-1}$ scanning speed.[95]

**Polyamide (PA)**

Polyamides (PAs) are semi-crystalline and thermoplastic aliphatic polymers, frequently referred to as Nylons, which contain recurring amide groups (R—C(=O)—NH—R') as integral parts of the main polymer chain. Synthetic PAs are typically produced by poly-condensation of diamines with dicarboxylic acids or esters. The aliphatic polyamides have been widely used as biomaterials for drug delivery or adhesives to be coupled to porous structures.[156] The relative high stiffness (i.e., E = 3 GPa),[157] the shape-holding features, the ease of processing, and the low biodegradability make them suitable for many biomedical applications, ranging from soft and flexible tubing and catheters to strong and stiff components for orthopedics and dental surgery.[158] Moreover, PAs have good biocompatibility with human tissues, probably due to their similarity to collagen protein in chemical structure and active group.[159]

Their high polarity gives to the PAs a relatively high affinity to form hydrogen bonds with nano-sized HA.[160] Nanoscale PA/HA scaffolds have been traditionally manufactured to reconstruct craniofacial, extremity and spinal column bone tissues via precipitation[161,162] or thermally induced phase inversion processing technique.[163] Porous n-HA/PA66 scaffolds were produced by thermal pressing and injection after preparing a mold by Xu *et al.* via SFF. They obtained 3D composites with pores size ranging between 300 – 500 μm, suitable for bone application, characterized by a very high CS (i.e., 117 MPa) and E of 5.6 GPa, features close to the actual bone.[164] They demonstrated that the new bone matrix was able to gradually creep into the interconnected porosity of HA/PA66 scaffolds, being embodied into the host tissue, without any fibrous tissue after 2 weeks, after both intramuscular and endosseous implantations in white rabbits (Figure 6).[97]

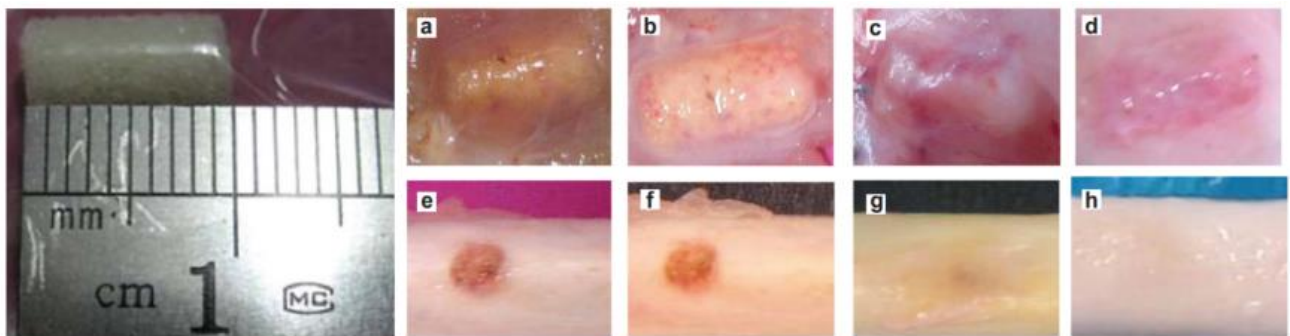

*Figure 6. PA66/HA scaffold dimensions (left). Observations at 2, 4, 12 and 26 weeks after (a-d) intramuscular and (e-h) endosseous implantation of PA66/HA scaffolds. Reprinted from* [97] *with permission – © 2011 Open Access, Dove Press.*

Li *et al.*, instead, exploited DIW to develop PA /nano-HA constructs to enhance the mandibular bone augmentation after a preliminary three-dimensional modelling of the anatomical structures supported by Computed Tomography (CT).[96]

**Poly(methyl methacrylate) (PMMA)**

Poly(methyl methacrylate) is an amorphous non-biodegradable thermoplastic material, approved by the FDA, used for reconstructive surgery applications such as *in situ* formed bone cement or pre-surgically shaped bone implant in the craniofacial area. PMMA is one of the amorphous polymers belonging to the acrylate family, which can be *in situ* polymerized from a slurry containing PMMA and methyl methacrylate monomers.[165] PMMA-based bone cements can be mixed with inorganic ceramics, or bioactive glass, to modulate curing kinetics and enforce mechanical properties. Commercially available PMMA-based bone cements, in the form of solid mold, are characterized by CSs of 85 – 100 MPa, which is close to cortical bone compressive strength ranging from 130 to 180 MPa.[166] However, the main drawbacks limiting their surgical application include: a high exothermic polymerization temperature, which can reach values as high as 40 – 56 °C,[167,168] a cement shrinkage of around 6 – 7% during the curing process *in vivo*,[167] presence of unreacted monomers, and relatively long operation times increasing the risk of infection and tissue necrosis.

Very little effort has been carried out on PMMA/HA composites, probably because of the non-biodegradability of the PMMA itself. Lee *et al.* investigated the use of SLS on a slurry composed of HA powder coated with PMMA and methanol,[99,100] achieving constructs with adequate porosity (i.e., 30% with pores size of 40 -100 µm). Although interesting, these papers did not report specific information on the mechanical properties of the scaffolds and/or any biological application/test to evaluate its performance with other tissues.

**Soft matrix-based biocomposites**

Soft matrix composites are mainly represented by hydrogel-based materials that are typically processed by extrusion-based AM technologies. Due to their nature, these soft-based composites are generally printed in a phase with a liquid-like predominant behavior. Subsequently, the printed HA-reinforced soft matrices usually undergo a post-processing step that allows the promotion of the long-term stability for the printed structures at physiological conditions. These procedures are mainly dependent on the printed material and great efforts have been made by several authors to identify the optimal crosslinking strategy for each material candidate (Figure 7) as described in detail in the following paragraphs.

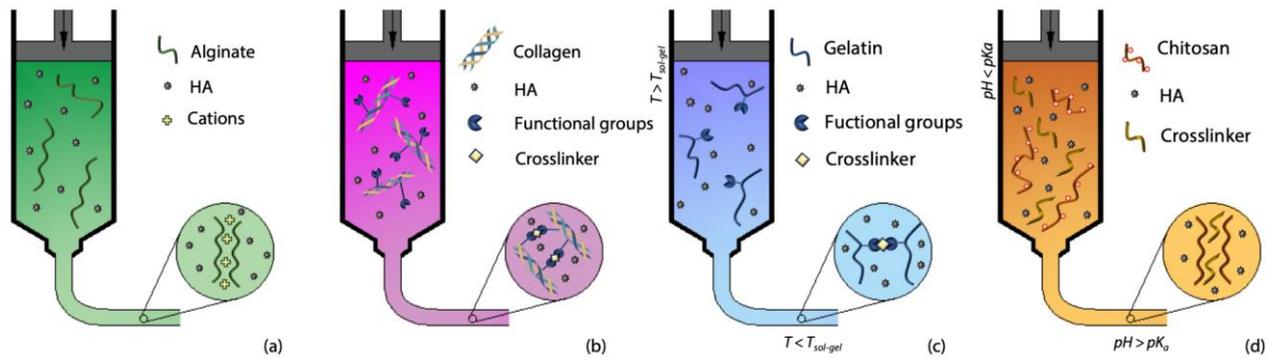

*Figure 7. Extrusion-based printing of HA-reinforced soft composites. The biomaterial ink is loaded in the cartridge of a 3D printer and printed to obtain the scaffold. Next, the printed structure is post-processed to stabilize the printed structure with a procedure optimized for the printed material. (a) Alginate is printed and, generally, immersed in a cationic bath to ionically crosslink its negatively charged groups. (b) Collagen is usually printed and then crosslinked through the activation of intermolecular bonds between aldehydes and other collagen amino acids with a crosslinker. (c) Gelatin is printed and, generally, chemically crosslinked by covalently binding its functional groups. (d) Chitosan is printed and, generally, immersed in an alkaline solution (pH > pK$_a$) to chemically crosslink its positively functional groups.*

The great interest in printing soft-matrix-HA-reinforced biomaterials relies on the possibility of simultaneously loading in the printed hydrogel, in addition to the HA reinforcement, cells and/or bioactive molecules to stimulate their proliferation/differentiation or a desired *in vivo* response (Table 5).

*Table 5. Soft matrix-based hydroxyapatite reinforced composites: structural, morphological and biological properties. For all the considered literature works, scaffolds are typically printed by layer-by-layer deposition of 0-90° shifted filaments. *HA-reinforced PVA was fabricated via SLS.*

| Material | Pore size (μm) | Printed filament dimension (μm) | Printed structure stabilization – crosslinking | Mechanical properties | HA size/distribution | Cell type | Cellular response | Reference |
|---|---|---|---|---|---|---|---|---|
| Alginate | > 500 | > 500 | CaCl$_2$ 200 mM | - | < 200 nm | Human Adipose-Derived Stem Cells (hASCs) | Viable cells successfully bioprinted. Increase expression of RUNX2, OCN and OSX osteogenic genes for HA-loads scaffolds vs. pure alginate | [104] |
| | > 500 | 100 - 400 | 2% w/v CaCl$_2$ | - | - | Mouse Embryo Cell (C3H) | Viable cells successfully bioprinted. Improved osteogenesis by Atsttrin release | [103] |
| | 500 – 1000 | > 500 | 2% w/v CaCl$_2$ | E = 30 – 40 kPa | - | hMSCs | Viable cells successfully bioprinted | [101] |
| | - | - | Na$_2$HPO$_4$ and CaSO$_4$ inside alginate ink | E = 2 – 10 kPa | - | MC3T3-E1 | Viable cells successfully bioprinted | [105] |
| | - | - | CaSO$_4$ internal gelation + 2% w/v CaCl$_2$ external gelation | 18 kPa | Uniform distribution of HA particles | MC3T3-E1 | Increased proliferation vs. pure alginate. Improved osteogenic gene expression and mineralization vs. pure alginate | [118] |
| | ≈ 500 | ≈ 500 | CaCl$_2$ 1 M | E = 6 – 10 kPa | HA distributed on the filament surface (34 µm thickness) | hMSCs | Increased attachment of cells on mineralized printed alginate vs. pure alginate | [102] |
| Collagen | 200 | - | Dehydrothermal process | E at low strains = 1.7 MPa E at high strains = 1.7 MPa | - | Osteosarcoma cell cells (MG-63) | Cell viability, proliferation, morphology, and distribution | [108] |
| Collagen | 800 | - | Dehydrothermal process | - | HA 1% w/v | - | - | [109] |
| Collagen | 71% | Up to 900 | 1% w/v genipin solution | E = 0.1 MPa | Uniform distribution of HA particles (diameter 100 nm) | BMSCs | Osteogenic proliferation and differentiation | [107] |
| Gelatin | ≈ 650 | ≈ 175 | Gelatin methacryloyl + LAP photoinitiatior | Shear store Modulus (G') ≈ 50 kPa Shear Loss Modulus (G'') ≈ 2 kPa | Spherical/plate/rod-like morphology 50 nm - 1 µm | hASCs from subcutaneous adipose tissue | Viable cells successfully bioprinted. Osteogenic differentiation proved by OPN and ALP staining | [111] |
| Gelatin | 375 ± 33 | 563 ± 32 | EDC solution | E = 0.4 MPa | Cementation of αTCP into calcium deficient HA | MG63 | Increased ALP activity and OCN expression for HA-loaded scaffolds vs. pure gelatin | [113] |
| Gelatin | 510 - 600 | 470 - 560 | Glutaraldehyde | CS = 2.7 – 3.8 MPa | Si-doped HA particles (≈ 35 nm) | MC3T3-E1 | Increased ALP activity and mineralized Extra Cellular Matrix (ECM) vs. pure gelatin. Proved antibacterial effect of | [112] |

| | | | | | | | |
|---|---|---|---|---|---|---|---|
| | | | | | | vancomycin-loaded scaffolds | |
| Chitosan | 200 - 400 | 150 | - | - | Particles 3 – 10 μm | Osteoblasts from Human Calvarial Bone Chips | Healthy morphology and strong proliferative activity | [114] |
| | - | - | - | E = 4.4 GPa | - | - | - | [116] |
| | 200 - 500 | 500 | - | - | nHA < 200 nm | MC3T3-E1 | Enhanced cell attachment, proliferation and differentiation | [117] |
| | 200 - 500 | - | - | - | nHA < 200 nm | MC3T3-E1 | High proliferation | [119] |
| | 100 | > 500 | Glycerol phosphate disodium salt | E = 14.97 MPa | nHA < 200 nm | MC3T3-E1 | Mineralization and differentiation | [118] |
| | 50% | - | - | E = 1.31 MPa/1.07 MPa | CH/HA 50:50 and CH/HA 60:40 | MC3T3-E1 | Excellent biocompatibility, proliferation | [72] |
| PVA* | - | - | - | - | Average diameter 60 μm / 70 w % with sprayed PVA | - | Bioactivity of the HA particles demonstrated by SBF | [120] |
| | - | - | - | - | Average diameter 60 μm / 10, 20, 30 wt% blended with PVA | - | Bioactivity of the HA particles demonstrated by SBF | [120] |
| Silk | 200 – 750 μm | 200 | - | Viscosity: $10^4$ Pas under low shear stresses and plateau at $10^5$ Pa E = 223±9 MPa | - | hMSCs | Good proliferation | [25] |
| | 400 μm / 70% | - | - | CS = 6 MPa | - | hBMSCs | Degradation of the construct. Cell proliferation and differentiation | [121] |

**Alginate**

Alginates constitute a family of polysaccharides naturally present in seaweeds or as bacterial exopolysaccharide. The alginate structure is composed by a linear repetition of 1→4 linked β-D-mannuronic acid (M) and α-L-guluronic acid (G) units, with $^4C_1$ ring conformation. The percentage and sequential distribution of G and M blocks (i.e., MMMM, GGGG or MGMG) determine alginate properties. Extraction of alginic acid from the selected natural source (typically seaweeds) is performed by removal of counterions, by immersion in mineral acid, and subsequent neutralization, by immersion in alkaline solution. Extracts are then filtered, washed and precipitated to obtain water-soluble sodium alginate.[169] Alginates form hydrogels in the presence of divalent cations, such as $Ca^{2+}$ and $Mg^{2+}$, due to the interaction of positive charges of ions with the negative charges on the alginate polymer chain leading to the so called "*egg-box*" structure.

Alginate hydrogels find extensive applications in the biomedical field, including drug delivery, protein delivery, wound dressings and tissue engineered scaffolds thanks to their proved biocompatibility.[170,171] The main drawbacks associated with the use of alginate-based hydrogels are the non-degradability in mammalians, the lack of intrinsically available cell-adhesive motifs (i.e., biological inertness) and their poor mechanical properties, typically < 500 kPa.[172] To overcome these limitations, oxidation of alginates was proposed by several authors to induce susceptibility to hydrolytic degradation.[173,174] Cell-adhesive motifs have been used to decorate alginate polymer chains by covalent binding of RGD peptides (Arginine-Glycine-Aspartate aminoacidic sequence) to promote the adhesion of cells by integrin-mediated binding and cell-matrix crosstalk.[175] Composite alginate hydrogels can be obtained by addition of micro/nano particles (e.g., iron oxide nanoparticles)[176] or fibers (e.g., PLA sub-micron fibers)[177] to the alginate hydrogel to improve its rheological properties and mechanical performance. Alternatively, the use of alginate-HA composites has been proposed to increase the hydrogel mechanical properties and improve the biomaterial-cells interaction, especially to target bone tissue regeneration. In particular, several authors have described the combination of alginate with HA to achieve ECM-biomimetic composites as scaffolds for bone regeneration; the addition of HA is performed to improve cell adhesion to the scaffold, its osteoconductive properties, and the hydrogel mechanical properties.[178] Moreover, HA reinforcement fosters the hydrogel radiopacity enhancing the visualization of implanted scaffolds by medical image analysis, including micro-CT.[104]

Alginates are among the most used hydrogels for the production of biomaterial inks and bioinks: in fact, mild crosslinking conditions, low costs, shear thinning properties, hydrophilicity, and fast gelation, which typically occurs in minutes, make alginate the optimal candidate for bioprinting processes.[179,180] Despite different AM technologies have been used for alginate processing, including droplet-based printing [181,182] and LAP,[183] alginate-based hydrogels and, in particular, alginate-HA composites are mainly processed by extrusion-based technologies.[184,185]

Alginate/HA composites can be customized to achieve rheological properties suitable for extrusion-based printing properties. Specifically, the alginate solution rheological properties can be increased by increasing the alginate concentration or by lowering the solution temperature.[186] The typical approach to print alginate composite filaments (i.e., diameter of 100 – 500 μm) is based on the extrusion of an alginate solution (i.e., typical concentrations 1.5 – 4% w/v) loaded with HA particles, in concentrations up to 20% w/v, and subsequent immersion in a divalent ion bath (i.e., typically 100 – 200 mM $CaCl_2$ solution). Alternatively[105] or complementarily to external gelation,[118] internal alginate gelation can be used by printing the alginate solution during its ongoing crosslinking process by adding $CaSO_4$ to the alginate ink before printing. Despite the intrinsic optimal properties of alginate solutions for extrusion-based printing, several studies have described the addition of gelatin to the bioink to further improve the printed shape maintenance immediately after printing.[104,106] In fact, HA-reinforced alginate/gelatin composites are largely printed on cooled substrates to promote temporary gelatin gelation ($T < T_{\text{sol-gel gelatin}}$) due to low temperatures and subsequently immersed in divalent ion bath to stabilize the shape for the long term by alginate crosslinking.[101] Alternatively, some authors[102] investigated the possibility of forming *in situ* HA after the immersion of $Na_2PO_4$-loaded alginate hydrogel in $CaCl_2$ solution: despite the HA formation having mainly been achieved on the outer shell of the printed filaments, this method represents an interesting alternative to the direct loading of HA into the printed ink (Figure 8).

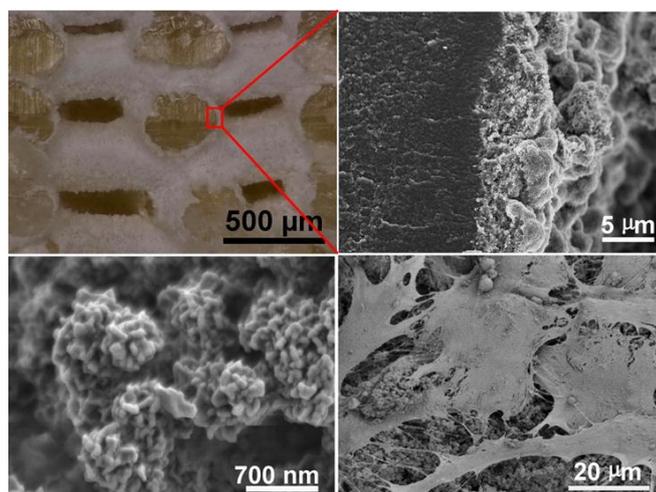

*Figure 8. In situ mineralized bioprinted alginate scaffolds. Nano-HA enucleated on the alginate printed filaments (stereomicroscopy images, top-left, and SEM micrographs, top-right and bottom-left). Bone marrow-derives mesenchymal stem cells adhered and spread on the printed scaffold (bottom-right). Reprinted from [102] with permission – © 2015 American Chemical Society.*

Despite several authors having reported an improvement of mechanical properties of alginate hydrogel scaffolds loaded with HA and fabricated by traditional methods, including freeze-drying,[178,187] only slight increases of the mechanical properties have been achieved, so far, for 3D-printed alginate-HA composites. In fact, an increase of the mechanical properties of printed alginate is generally observed, when loaded with HA, immediately after printing or in the dry state. However, after immersion in water or culture medium, the effect of HA addition on the mechanical properties is generally negligible: typical values for the compressive E are in the range of 10 – 30 kPa.[101,102,105] However, authors agree on the fact that improvements in the hydrogel mechanical properties can be expected after the deposition, by seeded cells, of inorganic ECM also promoted by HA constructs.[101] Improved osteogenesis of stem cells encapsulated in HA-reinforced alginate hydrogels has been proved in several works: HA is able to alter DNA methylation with the consequent modification of gene expression of an osteogenic phenotype. For instance, Wang *et al.* demonstrated that hASCs were characterized by the increased expression of osteogenesis-related genes (i.e., OCN and RUNX2) in nanoHA-loaded alginate hydrogels with respect to plain alginate hydrogels. Similarly, Demirtaş *et al.* proved an increase of osteogenic gene expression (i.e., OCN, OPN) of MC3T3-E1 cells loaded in alginate/HA bioinks compared to printed neat alginate.[118] By using a different approach, Luo *et al.* nucleated HA on the surface of 3D-printed alginate hydrogels by *in situ* mineralization [102] and observed an increased number of hBMSCs attached to the printed scaffolds, compared to printed alginate hydrogel without HA. Moreover, the presence of HA increased the intracellular levels of ALP activity, thus proving the increased functionality of the printed alginate when loaded with HA. The improved osteogenesis of HA-loaded 3D-printed alginate hydrogels, proved by *in vitro* tests, is also reflected in the osteoconductive properties of HA and ability in promoting the formation of new healthy bone once the scaffolds are implanted *in vivo*. For instance, alginate/HA biocomposites loaded with hASC were subcutaneously implanted *in vivo* in nude mice, after 7 days of osteogenic induction *in vitro*. The alginate/HA biocomposites were not only able to promote more bone tissue formation compared to plain alginate hydrogels, as proved by micro-CT imaging, but also to achieve bone formation throughout the whole printed construct, compared to pure alginate that promotes new bone formation only around the printed scaffolds pores.[104] To further improve the *in vivo* osteogenesis, alginate/HA biocomposite bioinks have been loaded with biomolecules/drugs that promote *in vivo* formation of bone tissue. For example, alginate/HA composite inks were loaded with Atsttrin, a Tumor Necrosis Factor (TNF)-α inhibitor, to reduce the inflammatory response after scaffold implantation. In this study, not only was the alginate/HA construct shown to improve bone formation in an *in vivo* calvarial mice defect, but the successful incorporation and subsequent release of Atsttrin from the printed scaffold was shown to modulate the local inflammatory response, thus further improving the bone healing process.[103]

**Collagen**

Collagen is a natural polymer widely present in the human body and, generally, in animals: from teeth to connective tissue, it represents the main component of numerous soft and hard tissues since it is the main constituent of the natural ECM.[188,189] Scientists have identified twenty-eight different polymers that, based on their structure, may be recognized as collagen. Collagen Type I, the most abundant, is one of the main components of the bone, being part of up to 89% of its organic matrix and 32% of its volumetric composition.[190] Collagen Type I also contains the RGD and Asp-Gly-Glu-Ala (DGEA) sequences that mediate cell binding via integrin receptors.[191–193] Other types of Collagen are Type II that has been found in cartilage as well as specific tissues of the auditory apparatus (i.e., the tympanic membrane) or Type III that composes the blood vessel walls.[47]

Collagen is a biodegradable and osteoconductive biomaterial[47,48] that provides natural attachment for cells.[194,195] Additionally, it represents an excellent natural carrier for bioactive molecules or drugs,[196,197] able to inhibit bacterial pathogens growth.[198] Despite its remarkable properties, collagen needs to be treated by crosslinking to improve the mechanical properties when hydrated,[199] specifically the stiffness (i.e., E in the order of 100 MPa),[200] that in most of the applications in tissue engineering is not adequate to bear the mechanical loads. On the other hand, besides improving E and resistance to enzymatic digestion,[201] chemically crosslinked collagen fibers can be potentially toxic due to residual molecules or compounds used for the crosslinking; thus; dehydration is generally the preferred alternative.[202]

Concerning bioengineering applications (e.g., tissue repairs and/or replacements), collagen derived from animals, especially bovines, is the most commonly used material due to its availability although it may elicit antigenic response.[203–205] Processing methods mainly include SFF to fabricate molds where to cast a HA-based blend, since its high viscosity makes difficult a direct 3D printing. In order to exploit the AM techniques, researchers have been employing chemical solvents to reduce the viscosity. However, since the effect of these solvents is still limited, the actual quantity of collagen in the biomaterial inks is limited.[107,206]

Collagen (typically Type I), is often reinforced with HA to achieve a biomimetic composite similar to bone. The reinforcement of the polymer through the HA ceramic structure enhances specific features for applications in surgery such as adaptability and shape control, clot facilitation and stabilization,[207] and higher degradation rate (i.e., 2 months vs. 2 years) with respect to the traditional polymeric scaffolds.[208] Additionally, porosity can be tailored by controlling the freezing rate, temperature and collagen concentration.[209–216] Being compatible with both humans and animals,[207,217] collagen/HA structures possess biological advantages compared to synthetic polymeric scaffolds.[218] Concerning bone regeneration, it has been demonstrated that both the composite constituents, singularly, are able to promote osteoblast differentiation but osteogenesis is accelerated when they are mixed.[219] Osteogenic cells appeared to better adhere, *in vitro*, to collagen surfaces instead of PLLAs and PGAs.[220] Furthermore, osteoconductivity is fostered if the composite is compared to monolithic HA.[217,221]

From a mechanical point of view, this biomaterial is able to balance the fragility of the HA with the ductile properties of the collagen allowing a better stability and resistance of the composite.[222,223] In order to evaluate the mechanical properties of the collagen/HA constructs, MD models have been coupled to experimental assessments in which researchers aimed at understanding the role played by HA content, morphology, porosity, pore architecture, and fabrication methods.[210,224–229] Among all the key features, Currey *et al.* demonstrated that high percentages of HA coupled with a reduced porosity of the scaffold may lead to higher values of E and ultimate strength.[230] At the same time, the collagen matrix acts as a load transfer medium to the rigid part (HA) deposited in its voids between tangled crosslinked fibers.[228,229] Collagen, hence, mechanically interacts with HA reinforcements by calcium ion bridges, leading to an increase of the composite resistance.[231] HA particles act as local stress concentrators in the collagen fibrous network and when the collagen fibers try to align on the direction of the stress, the material close to the HA particles gets a significant increase of the stress that can lead to fracture with a parallel reduction of the ultimate strength and final deformation.[194]

Different types of collagen/HA composites have been developed by scientists with diverse manufacturing approaches: dense[232–235] or porous materials,[215,221,224–226,236] or composites with elongated and plated-like HA crystals.[237] Different approaches have led to different results in terms of mechanical properties. For instance, a recent study employing HA whiskers to reinforce the SFF manufactured scaffold exhibited a nearly four-fold greater modulus compared to the equiaxial HA powder (i.e., HA 44 wt%). At the same time, no significant differences at higher and lower reinforcement levels were observed.[238]

The main challenge in fabricating collagen/HA scaffolds is related to the high viscosity of the composite that, as detailed above, can be partially overcome by using chemical solvents but, on the other hand, make difficult any inclusion of bioactive molecules[239]. These latter are, hence, mostly used to make surface coatings after the scaffold fabrication.[240,241] Due to these difficulties, the most pursued approach for producing collagen/HA structure is the indirect application of AM, namely SFF. In[108], the authors prepared a composite made of Collagen Type I and HA particles prepared with a freeze-drying procedure. After replacing the ice crystals with ethanol according to literature,[3,242] a dehydrothermal (DHT) crosslinking was performed. In this case, a slurry of material composed of 3% w/v suspension with equal parts in weight of collagen and HA microspheres in a water solution of 0.3 w/w acetic acid was cast in plastic molds to produce disk-shaped scaffolds. The constructs revealed larger pores (i.e., diameters up to 200 μm) than those found in unmineralized collagen scaffolds. From a mechanical point of view, the compressive modulus was 1.7 times, at low strains, and 2.8 times, at high strains, greater than that of collagen, reaching at most 50.74 kPa. After seeding cells, they showed a round shape, like healthy osteocytes, in the composite while in neat collagen structures they exhibited a more elongated geometry (Figure 9). Another attempt was carried out by Crystal et al. where collagen reinforced with HA flakes (i.e., 1% w/v) was cast into 3D-printed sacrificial molds post-degraded with ethanol, leading to an interconnected and branched porous structure with micro-channels up to 800 μm in width.[109]

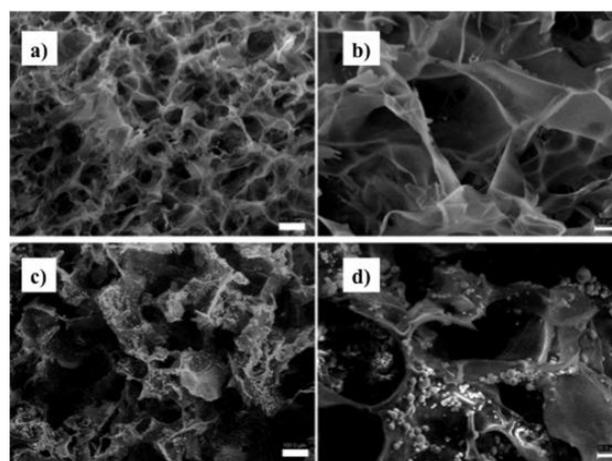

*Figure 9. SEM micrographs for (a, b) pure collagen and (c, d) collagen/HA scaffolds. Scale bars: (a,c) 100 μm, (b,d) 20 μm. Reprinted from [108] with permission – © 2016 Elsevier.*

A different approach was explored by Lin *et al.*, that used DIW to fabricate three-dimensional constructs at low temperature thanks to the recently-developed filament-free printing technique (Figure 10B-E).[107] With this technique, collagen and HA were mixed before printing (i.e., 1:2 w/w at 4 °C) in order to create structures with rods with a diameter range of 300 – 900 μm, where bioactive molecules could be included without any effect on their natural properties. In this case, the crosslinking was performed with 1% w/v genipin solution and sterilization was performed by ethylene oxide.[243–245] Although the prepared scaffold enhanced the BMSCs osteogenic proliferation and differentiation, the compressive modulus did not meet the value of the softest bone: 0.1 MPa vs. 2-20 MPa.[82]

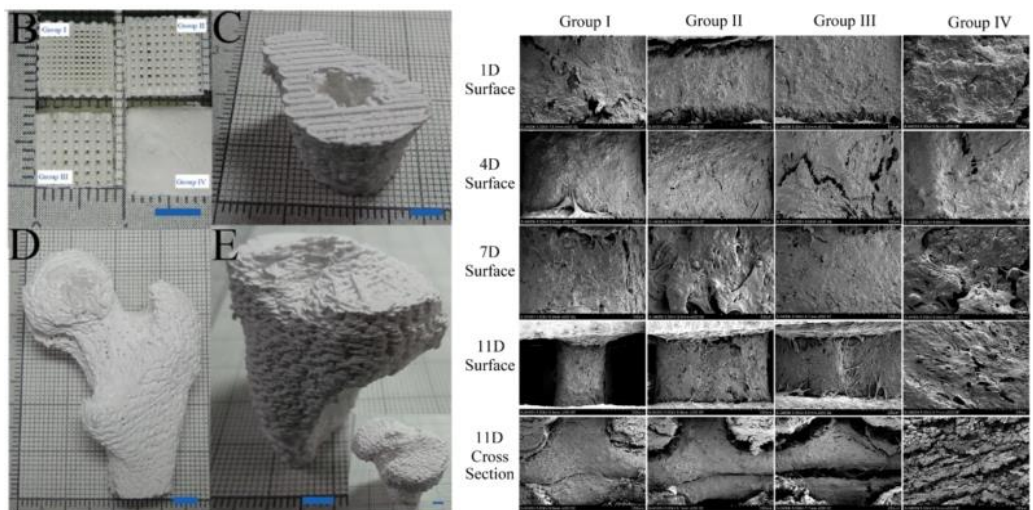

*Figure 10. (Left) Macroscopic view of the surface morphology of the experimental collagen/HA composite scaffolds. (B) Scaffolds had a microstructure with pore size of 400 μm and rod widths of 300 μm (group I), 600 μm (group II), or 900 μm (group III). Group IV was non-printed scaffolds. (C-E) Macrostructures mimicking human bones fabricated via DIW. Scale bar: 5 mm. (Right) SEM images of BMSCs on the surface of the scaffolds at 1, 4, 7, and 11 days after seeding in vitro showing cell growing in the constructs after 11 days from seeding. Reprinted from [107] with permission – © 2016 American Chemical Society.*

**Gelatin**

Gelatin is a water-soluble protein derived from the partial hydrolysis of collagen. Gelatin polymers with different $M_w$s and isoelectric points can be obtained from different animal tissues - typically porcine, bovine or fish – treated through an extraction process (i.e., alkaline or acidic collagen pre-treatments). The outstanding advantages in the use of gelatin as a naturally-derived polymer for tissue engineering applications rely in the exposure of cell-ligand motifs (i.e., RGD) which promote cells adhesion to gelatin by integrin-mediated interaction, and target sequences for metalloproteinases (Mitochondrial Processing Peptidase - MPP), which promote the *in vivo* enzymatic degradation and ECM remodeling. Moreover, compared to collagen, gelatin provokes less immunogenic and antigenic response after *in vivo* implantation. Furthermore, the popular use of gelatin for tissue engineering applications has been also due to its versatility, ease of availability and low cost.[246]

After dissolution in water, the gelatin solution is characterized by thermo-responsive properties. In fact, the gelatin solution undergoes a sol-gel transition at sol-gel temperature ($T_{sol-gel}$) $\approx$ 30 °C, depending on the specific gelatin properties and concentration; the solution is characterized by a liquid-like behavior at a temperature $T > T_{sol-gel}$, while it is characterized by a solid-like response for $T < T_{sol-gel}$. Thus, at 37 °C (i.e., *in vivo* temperature) the liquid-like response is predominant, not suitable for sustaining tissue regeneration by three-dimensional scaffold approach. Thus, crosslinking strategies must be used to improve the mechanical properties and stability of gelatin at $T > T_{sol-gel}$. Several approaches have been widely described including physical,[247] non-zero-[248] and zero-length[249] chemical, and enzymatic[250] crosslinking mechanisms that showed successful outcomes in generating biocompatible gelatin hydrogels for a variety of tissue engineering applications.

The abovementioned advantages of gelatin, together with its well-known thermo-responsive behavior, have made it one of the most popular materials for the production of ink and bioinks for extrusion-based processes. In fact, the thermo-responsive properties of gelatin can be smartly exploited by loading the gelatin solution in a printing cartridge ($T > T_{sol-gel}$) and printing on a cooled substrate ($T < T_{sol-gel}$) to temporarily fix the printed shape. Lastly, the printed construct is crosslinked to fix its long-term shape maintenance and improve stability at 37 °C. This printing strategy is by far the most used when considering additive manufacturing processes described to print gelatin/HA hydrogel inks.

Gelatin/HA biocomposites have been proposed for bone tissue engineering thanks to the joined advantages of the gelatin structure in mimicking the collagenous structure of bone and the inorganic HA particles that mimic the mineral bone component.[251] A few examples have described the use of such composites as biomaterial inks/bioinks by extrusion-based processing for the production of 3D-printed scaffolds (Figure 11).

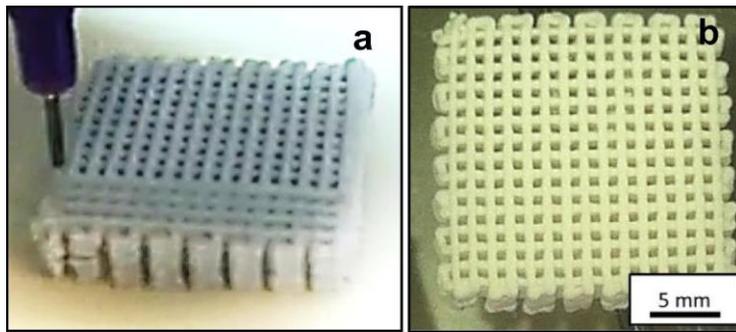

*Figure 11. 3D printing of HA-reinforced gelatin ink-based scaffold. Reprinted from [112] with permission – © 2015 Elsevier.*

The combined use of gelatin and loaded HA can recreate an ideal microenvironment for cells adhesion, proliferation and differentiation towards osteogenic phenotype, given by the presence of intrinsically cell-adhesive motifs of gelatin and the inorganic component represented by HA, together with improved mechanical properties given by the reinforcement of HA to sustain bone regeneration. Within the printing strategies, a controlled temperature for the extrusion-based process is generally required: in fact, literature works have described the printing of HA-loaded gelatin hydrogels by using a printing cartridge kept at T > $T_{sol\text{-}gel}$ and, generally, a printing plate with controlled temperature, T < $T_{sol\text{-}gel}$ since the thermo-responsive properties of gelatin hydrogels have been demonstrated to be not affected by the addition of HA.[113] Moreover, the addition of HA to the ink has been shown to improve the rheological response to shear stress by increasing the viscosity of the HA-loaded hydrogel; for instance, the principal rheological parameters (i.e., viscosity, *G'* and *G''*) were demonstrated to significantly increase after the addition of HA to the gelatin ink,[110] thus improving the shape definition of the printed scaffolds. The addition of a 2:1 ratio of HA to a 25% w/v gelatin solution more than doubles the complex viscosity *η\** of a gelatin ink and improves the shape definition of the printed solution compared to gelatin without HA, as shown by the printed filament circular cross-section obtained when printing the HA-loaded hydrogels compared to the collapsed filaments obtained by printing the neat gelatin solution.[113]

The addition of HA to the gelatin inks has been reported to influence not only its rheological properties, but also to improve the structural properties of the printed scaffolds with a generally proved increase of the mechanical properties comparing the HA-loaded hydrogels to the neat gelatin hydrogel. For instance, Huh *et al.*[113] demonstrated an increase of the compressive elastic modulus in gelatin hydrogels loaded with HA, printed and crosslinked by carbodiimide crosslinker - 1-ethyl-3-3-dimethylaminopropylcarbodiimide hydrochloride (EDC) (i.e., $E \approx 0.4$ MPa) when compared to the same hydrogel without HA (i.e., $E \approx 0.1$ MPa), both in the dry and wet state. Despite the obtained compressive modulus values being lower than that of spongy and compact bone (i.e., $E \approx 0.05 - 0.5$ and $14 - 20$ GPa, respectively)[251] even after the addition of HA, other authors demonstrated that the *in vitro* culture and osteogenic differentiation associated with cells inorganic ECM deposition increased the rheological properties of the printed scaffolds. Thus, even if the starting mechanical properties are lower than those of native bone, *in vitro* culture increase the structural properties of the scaffolds to target those of native bone, particularly significant when HA was loaded in the scaffolds before cell culture.[252]

In fact, authors working on HA-reinforced gelatin inks agree that, as demonstrated for other soft biocomposites (e.g., alginate), the addition of HA in the printed gelatin scaffold can promote cells osteogenesis, accomplished by osteogenic gene expression and inorganic matrix deposition. For instance, not only did HA-loaded gelatin hydrogels improve MG63 proliferation, which was possibly referred by the authors as a consequence of increased roughness obtained by addition of HA, but cells also showed increased ALP activity and OCN gene expression,[113] thus confirming the improved osteogenic effects of loaded HA. The observation made by using cell lines was then confirmed by more representative models using patient-derived ASCs, loaded in the gelatin-HA bioink, printed by micro-extrusion and crosslinked by UV curing (i.e., methacryloyl gelatin), which showed bone matrix production (i.e., ALP and OCN immunostaining) after 14 days of culture.

**Chitosan (CH)**

CH is a naturally derived polysaccharide obtained from the alkaline N-deacetylation of insoluble chitin. It is chemically composed of β-(1-4)-linked D-glucosamine and N-acetyl-D-glucosamine polysaccharide units.[253] The ratio between the two units is defined as degree of N-deacetylation. The presence of amino groups in the CH structure makes this polymer different from chitin, conferring it many peculiar properties. Indeed, for pH below its $pK_a$ (i.e., pH < 6.2), the amino groups ($NH_2$) on the CH chains are protonated into positively charged groups ($NH_3^+$) making it soluble.[254,255] By increasing pH, the amine groups become deprotonated to form insoluble CH polymer, which tends to produce a physical hydrogel thanks to reversible interactions (e.g., electrostatic, hydrophobic or hydrogen bonds) that can occur between polymer chains. This soluble-insoluble transition, at its $pK_a$ value, depends on the degree of N-deacetylation and $M_w$ of the polymer.[171,256] CH is considered as an appropriate functional material for biomedical applications due to its intrinsic properties: excellent biocompatibility, controlled biodegradability with safe by-products, antimicrobial and hemostatic properties.[254] The use of CH alone is mainly devoted to skin, nerves and soft tissue regeneration,[253,257] but the structural similarity of CH backbone to glycosaminoglycans, the main components of bone ECM, renders it able to support cell attachment and proliferation favoring chondrogenesis and bone tissue regeneration.[253,258] However, CH is a soft biomaterial characterized by low mechanical resistance, especially under hydrated conditions, which represents one of the main limitations in using it without the addition of other components. CH efficiently complexes metal ions or nanoparticles,[259] natural or synthetic anionic species (such as lipids, proteins, DNA), polyelectrolytes (such as tripolyphosphate), or is blended with other polymers[260] or functionalized with bioactive agents[261,262] to enhance its mechanical properties. This was also achieved by integrating bioceramics, in particular HA, into the CH matrix for scaffolding fabrication, showing that CH/HA scaffolds were characterized by a significant enhancement of mechanical strength with an increased osteoconductivity.[263–265] It has also been demonstrated that the addition of HA into CH scaffolds improved cell attachment, favoring a higher proliferation and a well-spread morphology when compared to the CH scaffolds alone.[266]

Within the recent years, CH has gained much attention for 3D scaffolds production with highly reproducible and controllable pore structure by AM due to its attractive properties, particularly its easy processability.[256,267] However, such AM approaches used for natural polymers require crosslinking treatment during or at completed printing process, because these water-soluble polymers are generally too soft to support their own structures after the printing process.[268] For this reason, most of the works reported in literature have been focused on the optimizing of CH-based hydrogel composition and rheological properties,[268,269] to make it a more easily printable polymer by extrusion-based approaches. For example, Lee *et al.* reported the production of highly porous CH scaffolds by extruding the chitosan solution onto a cryogenic plate held at -20 °C, and finally freeze-drying. They obtained a porous structure inside the filaments with relatively weak mechanical properties (i.e., E of 1.2 MPa and maximum tensile strength of 0.16 MPa for a dried scaffold).[270] Regarding the production of 3D CH-based scaffolds reinforced with HA by AM, a preliminary attempt was described by Ang *et al.* who fabricated CH/HA composites by using a rapid prototyping robotic dispensing (RPBOD) system where solutions of CH/HA were extruded through a small Teflon-lined nozzle (internal diameter 150 μm) into the dispensing medium (i.e., NaOH–ethanol) to form a CH gel-like precipitate.[114] 3D printed CH/HA scaffolds showed a good attachment between layers, forming a regular and reproducible macro-porous structure, fully interconnected, with pore size ranging between 200 – 400 μm. The high uniformity of the structure was likely due to the enhancement of mechanical strength given by the HA reinforcement, which allowed the scaffolds to hold their shape during the shrinkage phase in the dispensing medium. *In vitro* studies revealed a healthy morphology and a strong proliferative activity of osteoblasts seeded into the 3D-printed constructs. However, no data concerning the mechanical properties of the 3D-printed scaffolds were reported in the study and the main limitation of this technique was related to the formation of precipitated lumps in the nozzle. Additionally, this method suffers from the high sensitivity to CH concentration on the nozzle tip. During the process, the precipitation starts at the gel/dispersing medium interface immediately upon exposure, producing clots. The plotting CH-based material tended to solidify before it contacted base layer, resulting in poor adhesion and failure to hold the layers. To overcome these limitations, subsequently, Geng *et al.* improved the printing technique for the CH alone, by introducing a double nozzle system, allowing the sequential extrusion of CH and NaOH solution during the fabrication process.[115] The dual extrusion method eliminated the high sensitivity to material concentration compared to the previous work[114], because the precipitation occurs when the dispensing material and the coagulant medium merge on the base or on the previous layer. In this way, there were no precipitated lumps at the nozzle's end and no fluid medium movement to affect the shape of the precipitated layers of the scaffold.

Later, dense and porous cylindrical CH/HA scaffolds were fabricated by using an extrusion-based printer. The processing conditions involved the use of lactic acid (i.e., 40 wt%) as binder agent to different CH/HA composites (20 – 30 wt% of CH) followed by a post-hardening process, performed by incubating the printed scaffolds in NaOH and lactic acid solution at various concentrations, finally dried at room temperature. 3D-printed dense CH/HA (25 wt% of CH) showed optimal mechanical properties as demonstrated by their high compression strength of 16.32 MPa and Young Modulus of 4.4 GPa, with a very low porosity, about 37%. A compact layer of CS was observed for structures after immersion in 10 wt% lactic acid as binder. The collapse of the porous scaffold, observed during the post hardening process, was due to the immersion in a high concentration of solvent solution: therefore, the fabrication of 3D-printed CH/HA scaffolds for tissue engineering, whilst promising, still requires further optimization.[116]

Moreover, as the polymer coating on ceramics may hinder the exposure of the ceramics to the scaffold surfaces, the etching of scaffold surface could improve the hydrophilicity, the roughness and the surface chemistry of the scaffold, increasing the affinity of the composite towards cells. Thus, some studies have recently reported the production of CH scaffold reinforced with HA by AM coupled with surface plasma etching treatment. In particular, [117] fabricated 3D CH scaffolds containing 10 and 20% HA by using an air extrusion-based plotter. They showed a good porosity and interconnected structure, with a pore size ranging from 200 to 500 µm, while the increased hydrophilicity and bioactivity on the surface, exposed by plasma etching, enhanced MC3T3-E1 pre-osteoblast cell proliferation and differentiation. In particular, CH/HA (10% HA) scaffolds etched with $N_2$ plasma significantly improved MC3T3-E1 cell proliferation whilst CH/HA (20% HA) scaffolds etched with $O_2$ plasma showed the highest osteoblastic differentiation until 2 weeks (Figure 12).

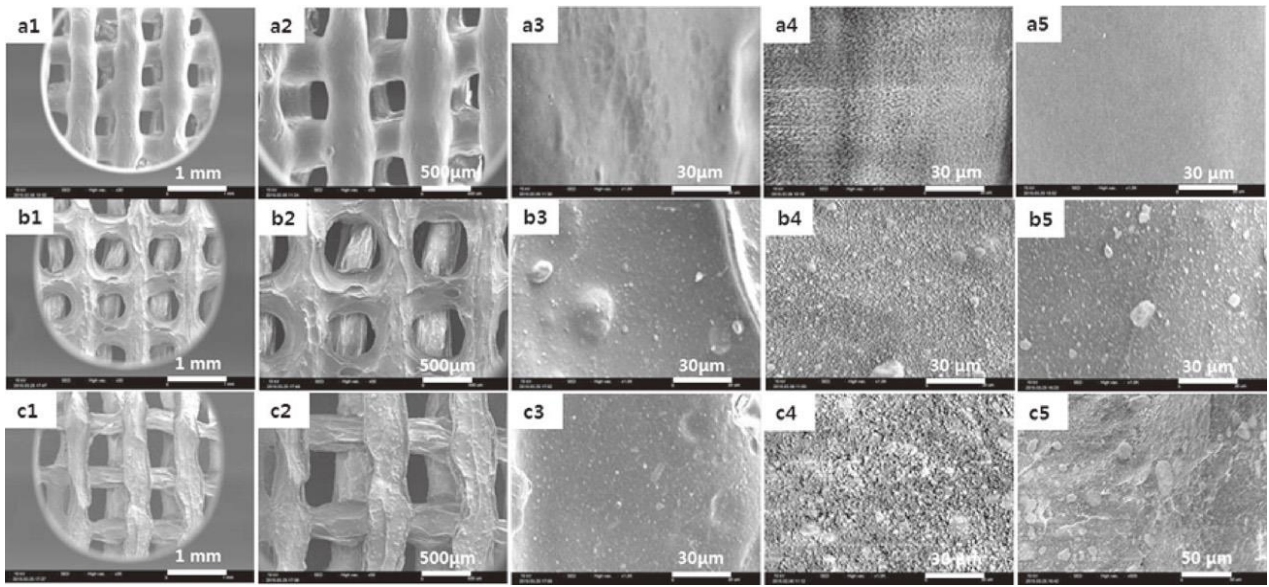

*Figure 12. SEM micrographs showing (a) CH (b) CH/HA (10 wt%) (c) CH/HA (20 wt%) scaffold after freeze-drying untreated and etched with $O_2$ and $N_2$ plasma, respectively. Reprinted from [117] with permission – © 2016 The Japan Society of Applied Physics.*

In 2017, a similar approach was reported by [119]. In this study, a CH/gelatin/HA scaffold with good interconnectivity and porosity was fabricated layer-by-layer by FDM and then its surfaces were etched by $O_2$ plasma to improve the roughness and wettability on the scaffold surface. The plasma-roughened surface enhanced the MC3T3-E1 pre-osteoblast cells' initial attachment but also their proliferation. However, in both cases, no data concerning the mechanical properties of the 3D-printed scaffolds were reported. Finally, a different approach was pursued by [72] where scaffolds made by SFF and composed of PLLA/CH/HA were compared to CH/HA scaffolds with different compositions (see PLA section).

Demirtaş *et al.* showed a first attempt of mixing chitosan and HA with cells and bioprinting a 3D scaffold. They mixed CH/HA solution (pH: 4.0) with glycerol phosphate disodium salt to generate printable scaffolds based on CH (pH: 6.95 – 7.0), at which MC3T3-E1 pre-osteoblast cells were added. Disk shaped hydrogels were printed with the use of an extruder-based bioprinter and held at 37 °C and 5% $CO_2$ for thermal ionic gelation. CH hydrogels exhibited a higher elastic modulus of 4.6 kPa that increased approximately 3-fold (i.e., 14.97 kPa) with the addition of HA. Morphologies of freeze-dried bioprinted CH revealed a porous structure (i.e., pores diameter of about 200 μm), which decreased to approximately 100 μm with the introduction of HA. It was also observed that cells printed within CH/HA composite hydrogels were homogeneously distributed inside of the structures and showed peak expression levels for early and late stages osteogenic markers particularly in the presence of HA.[118]

**Polyvinyl alcohol (PVA)**

Polyvinyl alcohol (PVA) is a water-soluble synthetic polymer, with a linear structure, derived from partial or full polyvinyl acetate hydroxylation. The hydroxylation process affects its physical, chemical and mechanical properties. In fact, the degree of hydroxylation influences PVA molecular weight and solubility, and consequently its swelling behavior and E. A higher degree of hydroxylation and polymerization of the PVA induces a lower solubility in water. For this reason, its water solubility makes necessary the use of physical or chemical crosslinking agents (e.g. gamma irradiation, glutaraldehyde, genipin and others), for granting structural stability and making PVA easier to process as a hydrogel, able to swell in the presence of biological fluids.[271] PVA has been approved by the FDA for food packaging, because of its excellent barrier properties when processed in the form of films. However, it has been widely used also as a biomaterial for medical device fabrication, due to its biocompatibility, non-toxicity, non-carcinogenicity, swelling properties, and bio-adhesive characteristics.[272] In fact, it has been extensively proposed as a replacement for cartilage and meniscus defects due to its high water content, and rubber elastic physical properties, which make PVA a good candidate for load-bearing applications.[271] More specifically, PVA hydrogels have exhibited a tensile strength in the range of 1 - 17 MPa[273] and a compressive modulus ranging from 0.0012 to 0.85 MPa, which is close to cartilaginous tissue mechanical features (i.e., tensile strength of 17 MPa[274] and compressive modulus varying between 0.53 and 1.82 MPa).[275]

Being processed at temperature used on the SLS, several groups have investigated the use of PVA to fabricate porous scaffolds for bone tissue engineering by AM. In[120], the authors approached the fabrication of PVA/HA constructs by using SLS on two different raw materials: PVA coated via spray drying technique with HA powder (70 wt%) and a slurry of PVA and HA powder – diameter size less than 60 μm – with different HA concentrations (%HA varied: 10, 20, 30 wt%). The conclusions of this study highlighted how the SLS-machined-blended composite presented the highest porosity with a good grade of interconnectivity. Biological tests showed also that the laser employment did not affect the bioactivity of the HA in both cases, although the blended mix has to be preferred for tissue engineering applications due to its higher compatibility with the host tissues, as demonstrated by tests in Simulated Body Fluid (SBF).

**Silk**

Silk is a natural protein fiber possessing high mechanical strength, tunability, controllable degradation and manufacturing flexibility as well as good biocompatibility.[276–285] Scientists have been using this material for a number of applications including medical sutures, and tissue regeneration since it was observed that sericin has some inflammatory effects.[286–293] Despite silk, considering its high mechanical properties, can be used as reinforcement to improve the features of soft matrices, it can also be used for the production of soft composites and can thus be considered a soft material.

Although a number a studies have been carried out to demonstrate how silk fibroin alone can be 3D-printed through inkjet printing,[294] very few studies have been carried out on silk-HA biocomposites. In [25], the authors used Direct Ink Writing (DIW) to fabricate three-dimensional silk fibroin-HA structures aimed at regenerating bone tissues. They fabricated structures (i.e., filament diameter = 200 μm) with pores ranging from 200 to 750 μm able to promote osteogenesis and vasculogenesis. The high-concentrated HA biomaterial ink, deposited at 2 mm·s$^{-1}$, possessed high viscosity (i.e., $10^4$ Pa·s under low shear stresses and plateau at $10^5$ Pa). Mechanical and biological assessments showed interesting properties including a stiffness of 220 MPa for single filaments and a relevant promotion and proliferation of hMSCs seeded for bone regeneration (Figure 13).

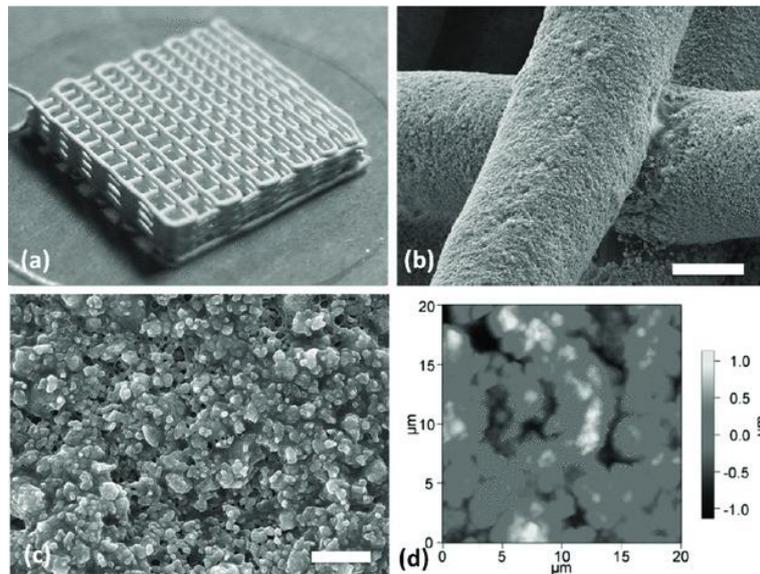

*Figure 13. (a) 3D silk/HA scaffold with gradient porosity. (b) SEM micrograph of printed silk/HA filaments: detail of an overlapping of two layers. Scale bar, 100 μm. (c) SEM micrograph of the of the silk/HA filament surface. Scale bar 10 μm. (d) Height profile of a representative silk/HA filament acquired by AFM. Reprinted from [25] with permission – © 2012 John Wiley and Sons.*

More recently, silk/HA composites were studied by Huang *et al.* in combination with sodium alginate with different mass ratios (i.e., 3:1, 5:1 and 10:1), fabricating via DIW porous cylindrical scaffolds (porosity = 70% - pores size = 400 μm) that revealed CSs in the order of 6 MPa. The constructs, tested with hBMSCs, showed good biological properties for bone regeneration purposes (i.e., cell adhesion and penetration, proliferation and osteogenic differentiation).[121]

**Conclusions and future perspectives**

Regeneration of human tissues by tissue engineering approach is still a very challenging goal since it requires the accurate design of scaffolds that sustain the regeneration of the damaged tissue by balancing the scaffolds properties, including (i) an adequate structural support and biomimetic mechanical properties, (ii) an optimal porosity to allow cell colonization and tissue infiltration, and (iii) cyto and biocompatibility. Additive manufacturing techniques have gained tremendous interest and success, especially within the last two decades, due to their versatility over fabricating devices from the micro to the macro scale, imposing their advantages (e.g., cost-effectiveness, relative inexpensiveness) on the traditional manufacturing processes (e.g., machining).

Besides the numerous possibilities offered by AM, the fabrication of structures made of a desired material, often in combination with other materials (i.e., composites), can represent a challenge due to particular requirements and features possessed either by the materials themselves or the process itself. This is particularly true in the specific case of HA-reinforced composites, that are used mainly as tools to promote/replace cartilaginous and bone structures due to their chemical similarities to the native tissues (Figure 14). Besides the matrix used to embody HA, the addition of this hard component induces a restriction on the AM options. Specifically, the use of the droplet-based technique has not been implemented and all the other approaches, depicted schematically in Figure 2, have suffered the limitations caused by the high viscosity of the ink (i.e., hard-matrix-based composites), or the poor processability by laser-based techniques (i.e., soft-matrix-based materials). This is why, as also summarized in Table 2, there is almost clear division among the HA-reinforced composites with respect to the AM techniques.

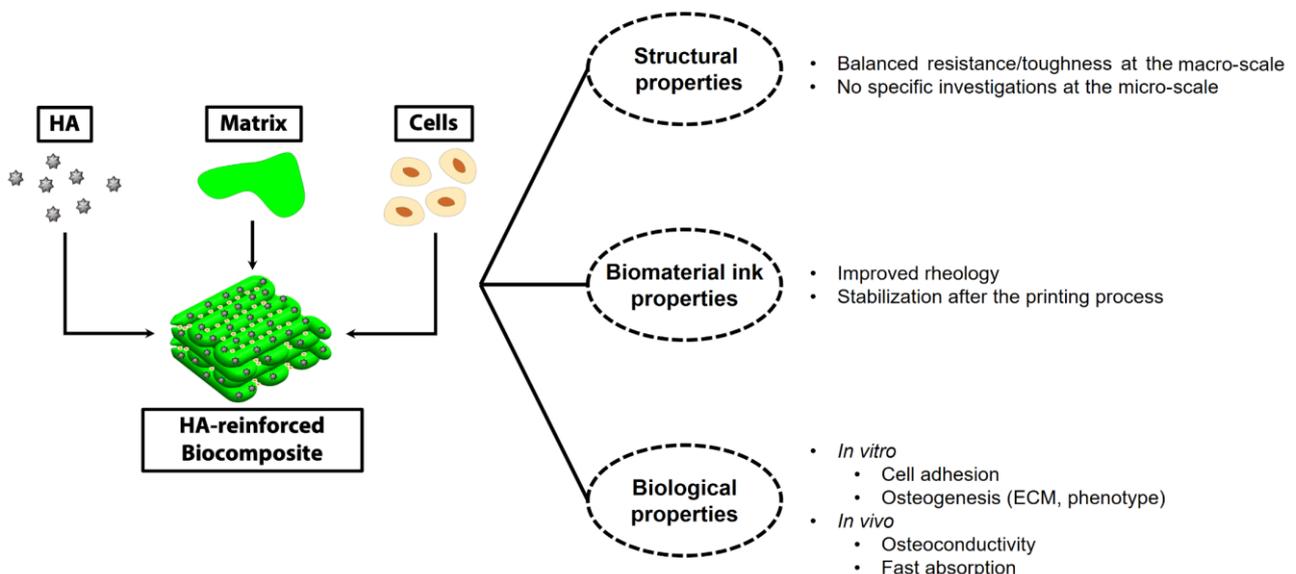

*Figure 14. Deployment of the main features possessed by polymeric scaffolds when reinforced with HA.*

Moreover, by considering the composition of the HA-reinforced structures, it appears evident how the only class of matrices that has been processed so far with AM is the polymeric one. The main reason may be found in the difficulties in managing metallic-/ceramic-matrix-based materials that can be difficult to process due to their poor toughness and high viscosity.

Polymeric hard-matrix-based composites have been generally preferred as load-bearing replacements, due to their higher mechanical properties, in contrast to the soft-matrix-based ones that have been mainly proposed for non-load-bearing applications, if not properly crosslinked. However, recent trends have shown much more interest for the soft materials due to their easier processability and shapeability in addition to the possibility to directly load and print viable cells in the compounds, to obtain a uniform cell distribution in the printed filaments. Soft materials do, hence, represent the most relevant research topic for scientists that, as detailed above, are still developing new strategies to overcome their main drawback, namely the poorer mechanical properties (vs. thermoplastic polymers), through different crosslinking strategies.

A detailed analysis and achievement of satisfying mechanical properties still represent, however, a frontier: while hard materials may apparently represent a good solution to bear loads, a deep study on understanding how each parameter (e.g., %porosity, pores size) acts, especially at the microscale, is still missing. Moreover, it remains still unclear how the HA and its features (e.g., size, distribution) affect the mechanical properties at cell level. A tool to investigate these issues may be represented by an *in silico* approach where computational modeling at multiple scales could be the key to unveil the mechanical behavior of the material, as already demonstrated for the specific case of silk.[281,282,295]

Another feature little investigated is the vascularization of the HA-reinforced scaffolds: it is clear that an interconnected porosity represents a necessary condition for vascularization, but few studies have evaluated the capability of the constructs to allow the flows of nutrients needed to avoid the necrosis of the tissues and the embodiment in the host site.[296–298]

From a biological point of view, many degradation studies have been carried out that, however, have not achieved a precise control of the phenomenon. A balanced degradation of the scaffold is fundamental, indeed, for a correct regeneration of the tissues, reducing the inflammatory response and granting an adequate mechanical stability throughout the process of tissue regeneration. Furthermore, many studies showed how the development of such HA-reinforced tissues is supported by successful *in vitro* tests. However, biological tests are generally still missing *in vivo* validation, whose results may confirm, or not, the laboratory tests.

As broadly discussed, the scientific community has mainly pursued the goal to realize a biocompatible load-bearing bone-mimicking construct to be able to be properly hosted by native tissues. However, we envision the possibility of a more complex synergistic approach that could deliver the next generation bone tissue in which the construct itself will be able, through its topology and physico-chemical properties, to overcome the vascularization issues that, still represent the toughest challenge for scientists. This ambitious goal could be progressively achieved in the near future by a complementary approach involving both modeling and experiments. Specifically, multiscale modeling from molecular to continuum scale, supported by experimental results, will gradually unveil the physico-chemical mechanisms still unknown reaching a complete understanding of the phenomena overcoming the above described issues.

In addition to the scientific aspects, some translational and ethical issues must be considered in the very near future. Developing new scientific and technological products has become really challenging due to the high research costs coupled to the increased reduction of external funding and grants. *In vitro* tests, although successful, cannot be the last step in this research field: the lack of the aforementioned *in vivo* studies is mainly due to the increased high costs to be faced when the experimentation is moving towards human clinical trials.

From a legislative point of view, due to rapid development of the AM technologies for biomedical constructs, there is an increasing need for a robust national/international regulation on how to develop, sterilize and promote the use of these new engineered materials.[299] Recently, the FDA has released specific guidelines on technical considerations for AM devices describing how the whole process (from the design to the post-processing control) has to comply with specific criteria in order to receive the approval for clinical uses.[300] Ideally, after receiving the FDA certification, each hospital could possess a dedicated center able to collect the requests from all the departments to manufacture bio-constructs for the specific requirements of each patient, including its own cells.

Additionally, it would help disseminating the advantages of such an approach, overcoming also any mistrust that is often boosted by unreferenced sources.

Table 6 summarizes a selection of case studies that showcase the potential impact of AM technologies for tissue engineering applications.

*Table 6. Selected case studies for AM HA-based constructs.*

| AM method | Matrix | Application | Reference |
|---|---|---|---|
| DIW | PA | Reconstruction of condylar defect | [96] |
| DIW | PCL | Human molars and rat incisors | [87] |
| DIW | PA | Mandibular bone augmentation | [96] |
| DIW | Alginate | Calvarian mice defect | [103] |
| FDM | PLLA/quaternarized CH | Cortical bone reconstruction in rats and cancellous bone reconstruction in rabbits | [85] |
| SFF | PA66 | Intramuscular and endosseous implantation | [97] |

In conclusion, despite the encouraging results achieved for both the hard/soft materials there are still several questions that need to be answered in order to properly tune the HA-based materials for each specific purpose and fabrication process. A better understanding on the sensitivity of the final outcome for each involved parameter, of the material and of the AM process, will definitely represent the key to fabricate improved customized devices for replacing/regenerating cartilaginous/bone tissues with strong implications also on other classes of biocomposites processed via AM in view of other applications in the biomedical field.

**Conflict of interest**

All the authors declare no conflict of interest.

**Acknowledgments**

This work has received funding from the European Union's Horizon 2020 research and innovation program under the Marie Skłodowska-Curie grant agreement COLLHEAR No 794614 who supported MM. SS was financially supported by RECOVER project, funded by the Ministry of Agriculture, Forestry and Tourism (MIPAAF), call 3549, Project no. CUPJ57G17000150008. NCN was financially supported by a Progetto Roberta Rocca (Doctoral at MIT) Fellowship.